\documentclass[12pt]{article}

\usepackage[utf8]{inputenc}
\usepackage{amsmath}
\usepackage{amssymb}
\usepackage{graphicx}
\usepackage{grffile}
\usepackage{color}
\usepackage{booktabs}
\usepackage{subfigure}
\usepackage{rotating}
\usepackage{multicol}
\usepackage{multirow}
\usepackage{sidecap}

\hypersetup{ 
  colorlinks=true, 
  pdftitle={Modeling the Evolution of Networks as Shrinking Structural Diversity},
  pdfauthor={Jérôme Kunegis}, 
  citecolor=black, urlcolor=black, linkcolor=black
}

\newcommand{\wTwo}{0.47\textwidth}
\newcommand{\wTwoPointFive}{0.38\textwidth}

\begin{document}

\title{
  Modeling the Evolution of Networks as Shrinking Structural Diversity 
}

\author{
  Jérôme Kunegis \\
  University of Namur
}

\maketitle

\begin{abstract}
  This article reviews and evaluates models of network evolution based
  on the notion of structural diversity.  We show that diversity is an
  underlying theme of three principles of network evolution: the
  preferential attachment model, connectivity and link prediction.  We
  show that in all three cases, a dominant trend towards shrinking
  diversity is apparent, both theoretically and empirically.  In
  previous work, many kinds of different data have been modeled as
  networks: social structure, navigational structure, transport
  infrastructure, communication, etc.  Almost all these types of
  networks are not static structures, but instead dynamic systems that
  change continuously. Thus, an important question concerns the trends
  observable in these networks and their interpretation in terms of
  existing network models.  We show in this article that most numerical
  network characteristics follow statistically significant trends going
  either up or down, and that these trends can be predicted by
  considering the notion of diversity.  Our work extends previous work
  observing a shrinking network diameter to measures such as the
  clustering coefficient, power-law exponent and random walk return
  probability, and justifies preferential attachment models and link
  prediction algorithms.  We evaluate our hypothesis experimentally
  using a diverse collection of twenty-seven temporally evolving
  real-world network datasets.
\end{abstract}

\section{Introduction}
Networks provide an adequate and established model for studying a broad
range of complex structures and processes.  Common examples include
friendships between people, Web pages connected by hyperlinks,
interactions between users and content items, references between
scientific works, communication and transport infrastructure. What these
examples all have in common is their dynamics: Over time, all networks
change. In some cases, new edges appear, such as in an email network
when a new email is sent, or new nodes appear, such as in a scientific
publication network in which a new author appears. An important question
in the area of network analysis is thus: What are the patterns under
which networks evolve?  This question can and has been answered in many
different ways, depending on the type of network and the network measure
considered.  In this work, we argue that the question can be answered
by the notion of diversity.  We note that many numerical network
measures can be interpreted 
as a measure of structural diversity, and are observed in actual
networks to evolve 
monotonically over time. The main observation that we make is that
these numerical measures are as a general rule evolving in the direction
that can be interpreted as shrinking diversity, 
regardless of the application area of the network. 
This statement is of course not an absolute one~--
it is not true that \emph{all} network measures evolve consistently
towards less diversity over time in \emph{all} network types. Instead,
we will show that this happens in a high number of cases, which is
statistically and systematically larger than what would be expected by a
random behavior.

The notion of diversity that we consider is entirely structural.  In
other words, we consider only the structure of a network, and not its
content. As an example, we do not consider the change in the contents of
emails written by people over time, but only \emph{who} people are
writing to.
However, content can often be
represented as a network, and in that
case we do consider it.  For 
instance, the bag of words model can be seen as a bipartite graph
connecting documents with words. 
To make the notion of diversity more precise, we propose
three principles of network evolution that fall under the umbrella term
\emph{shrinking diversity}: 
\begin{itemize}
\item \textbf{Preferential attachment}:  The notion that \emph{the
  richer get richer}, i.e., that nodes with many
  neighbors tend to attract new links faster than other nodes is a
  well-established principle in modeling networks, and constitutes the
  basis for the scale-free graph model of Barabási and
  Albert~\cite{b439} and many other models. 
  Measures of the equality of the degree distribution can be
  interpreted as measures of diversity and are
  predicted to increase under this model. 
\item \textbf{Increasing connectivity}:  Many numerical network
  characteristics can be interpreted as a form of
  connectivity. In a highly connected network, less subgraphs with low
  connectivity to the rest of the network exist, and thus an increasing
  connectivity 
  can be interpreted as shrinking diversity. 
\item \textbf{Link prediction algorithms}:  The problem of link
  prediction is the task of predicting which edges 
  will appear in a network, given the current network. For a given link
  prediction algorithm, we 
  can identify numerical network measures that must increase
  or decrease over time if edges are added according to the predictions
  of the algorithm. A network that grows according to a link prediction 
  algorithm will tend to consolidate its structure and not add any new
  structure, decreasing the diversity of the network. 
\end{itemize}

The contribution of this work is thus to (1)~review numerical network
measures in light of the notion of structural diversity, (2)~introduce a
methodology 
for measuring whether a numerical network measure is significantly
increasing or decreasing, (3)~perform corresponding tests for a large
set of numerical measures and network datasets from different areas, and
(4)~show that the
three network evolution principles are valid, validating the shrinking
diversity hypothesis. 
The article is structured as follows:
Section~\ref{sec:diversity} gives the
  definition of the concept of diversity, 
  states the three principles of network evolution, 
  and states our hypothesis of shrinking diversity. 
  Section~\ref{sec:measures} introduces the eleven network
  measures we consider. 
  Section~\ref{sec:experiments} describes our systematic
  experiments on a 
  collection of twenty-seven temporal network datasets in order to test
  our hypothesis. 
We conclude in Section~\ref{sec:conclusion}. 
This article is partially based on previous conference papers by the
author~\cite{kunegis:power-law,kunegis:network-rank}. 

\section{Network Evolution and Diversity}
\label{sec:diversity}
Diversity is generally defined as the quality of a collection of things
containing many 
\emph{different} or \emph{unlike} objects. In the context of networks,
the diversity of a system can be understood as the diversity of 
opinions, topics, communities or any other entities represented by the
network. For instance, in a movie recommender 
system, we understand that the community has more diversity when the
movies being watched and rated are different from one user to another.
On the other 
hand, a community in which most people watch the same fixed set of
movies is not diverse.  This notion of diversity is
independent of the notion of size:  A movie recommender community may
have many users and include many movies, and still lack diversity, because
most users have seen the same set of movies.  Thus, diversity does not
denote the size but the distinctness of the content. In the context of a
network such as the user--movie graph, diversity is thus achieved when
many users have seen \emph{different} sets of movies. Equivalently, we
can require that individual movies have been seen by different sets of
users.  

\begin{table}
  \caption{
    \label{tab:comparison}
    The measures of diversity we study.  The first column gives the aspect
    of a network that is covered by the measure.  The second column
    describes in what case a network can be called diverse under that
    aspect.  
  }
  \centering
  \begin{tabular}{ l l p{6.6cm} }
    \toprule
    & \textbf{Aspect} & \textbf{A network has shrinking diversity when} \\  
    \midrule
    \ref{sec:richer} & Preferential attachment & 
    Nodes with many neighbors acquire new neighbors faster than others \\
    \ref{sec:connect} & Connectivity & 
    Distinct parts of the network become better connected over time \\
    \ref{sec:linkpre} & Link prediction & 
    Its evolution follows link prediction functions \\
    \bottomrule
  \end{tabular}
\end{table}

We investigate three concepts of network analysis that capture the
notion of diversity: 
\begin{itemize}
\item \textbf{Preferential attachment}:  
  By the principle of preferential attachment, nodes with high degree
  receive new edges faster than nodes with small degree, implying that
  the inequality of the degree distribution increases over time. 
  Thus, a network has
  diversity when all nodes have approximately the same number of
  neighbors, and has low diversity when some nodes have many
  neighbors and some nodes have very few neighbors.
\item \textbf{Increasing connectivity}: 
  A well-known result in network analysis is that of a shrinking
  diameter, i.e., the observation that the diameter of real-world
  networks tends to decrease over time.  This result can be generalized
  to other measures of connectivity, giving predictions of increasing
  connectivity, which can be interpreted as shrinking diversity, in the
  sense that a network that is well connected allows less room for
  individual communities, and thus is less diverse. 
\item \textbf{Link prediction functions}:  A network has shrinking diversity
  when its evolution follows link prediction functions. A function
  that predicts the evolution of a network cannot predict new 
  network structures (since they do not exist yet), but can only predict
  the strengthening of existing network structure. Thus, if a network
  evolves in accordance with a link prediction function, we interpret
  its diversity as shrinking. The numerical measures in this category
  are thus such that their temporal evolution is monotonous under
  various link prediction models. 
\end{itemize}
Based on these three aspects of diversity, we derive individual
numerical measures that we interpret in terms of diversity in
Section~\ref{sec:measures}. 
Table~\ref{tab:comparison} gives an overview of the aspects and
measures. 

\subsection{Related Work}
The concept of \emph{diversity} is broad and relates to many different areas of
research, with many different definitions.  In this subsection, we very briefly
mention approaches to the concept that are not directly used in the rest of the work. 

Changes in variables describing communities have been studied in organizational behavior studies \cite{aldrich1976}.  
An article by Hannan \cite{hannan1977} analyses stability of organizational structures, in
the continuous domain, and on the level of individual organizations.
Another article \cite{hannan1984} considers the inertia of organizations.  A study about
multiple organizations and their decline (which can be interpreted as
less diversity) is given by \cite{hannan1988}.  

Similarly, the competitive exclusion principle \cite{hardin1960} can be
interpreted as a decrease in diversity:  the fact that multiple species
competing in a single ecological niche does not represent a stable
equilibrium, but will tend to the extinction of all but one species.  

In sociology, inequalities in ``human'' distributions have been considered, as
well as urban
hierarchies, and how humans adapt,~\cite{hawley1981}.
In this context, humans tend to move to an equilibrium, or
evolve~\cite{hawley1986}. 

The concept of the diffusion of innovations by Rogers
\cite{diffusion-of-innovations} is an early (pre-network) theory that
explains the diffusion of an innovation.  It can be seen as related to
the concept of preferential attachment in that people are likely to
connected to new, popular trends (innovations).  In terms of diversity,
the adoption of a single innovation by a population can be interpreted
as the reduction of diversity, as it results in the complete population
using one and the same technology. 

Shrinking diversity is also explained in sociology by the theory of
ingroups and outgroups \cite{group-bias,ingroups-outgroups}, in which
people tend to favor persons from their own social groups, thus
reinforcing existing social structures. 

Also related are theories of sociological change~\cite{monge2003}. 

The article \cite{monge2008} reviews multiple models of (among others) changes in
network properties.
The review covers various densities, Leskovec's
diameter, evolutionary-like approaches, whose outcome
depends strongly on the community's properties, and thus do not
generalize.  The work also considers communities that interact with each other,
explaining the structural changes in each community by their interaction
with other communities.  

Homophily can also explain the dynamics of groups, i.e., that people
tend to form groups with other persons that have similar attributes to
themselves, resulting in clustered social networks \cite{hinds2000}. 

Another related field of research is that of the dynamics of
interorganizational ties \cite{inter3,inter1,inter2}, according to which organizational
systems reach capacity in terms of the number of competitors that are in
a field.  When a field reaches capacity, the nature of the relationships
changes from competitors helping and communicating with each other to
more conventionally competivity behavior and turning to network ties
that are more complementary, thus decreasing the diversity of the
network. 

Case studies that have observed shrinking diversity include the work by Bryant
and Monge~\cite{children-television}, as well as that by Brown and Ashman
\cite{intersectoral}.   

\subsection{Evolution of Diversity Measures}
Within the field of network analysis, an important aspect is that of
network evolution, i.e., the understanding of the temporal changes in a
network's structure. 
A well-known example of a temporal trend in a numerical network measure is that
of a shrinking diameter.  By studying several large temporal networks,
Leskovec et al.~\cite{b242} discovered that the 90-percentile effective diameter
(i.e., the average number of hops needed to reach 90\% of a network's
nodes) shrinks over time. This result leads to several
follow-up questions: Since the diameter can be understood as a measure
of connectivity, are other measures of connectivity shrinking too?
Since the authors of that work described a network growth model to
explain shrinking diameters, do other network growth models also lead to
monotonous trends in network measures?  In order to answer this and
related questions, this article will study eleven numerical network
measures.  The hypothesis studied in this paper is as follows: 

\textbf{Hypothesis}  \textit{The structural diversity of most evolving networks
is shrinking.}

During the lifetime of a network, a given network measure will be subject
to nontrivial fluctuations, and the temporal evolution of network
measures is not always necessarily monotonous.  For instance, the
network density (i.e., the 
mean degree of nodes) has been shown to first grow very fast, then
decline, and then end up growing slowly for the rest of a network's
lifetime~\cite{b700}.  
As this example shows, it is important to distinguish between the
behavior of a network measure of the network's full lifetime, and the
behavior of a network measure at one specific point in the lifetime of
the network. Thus, we will test both cases in our experiments. 

Since it is not possible to 
define in a general manner what \emph{most} networks represent, we are
only able to test this hypothesis on a large collection of networks that
are available to us. 
Although these networks are from a diverse range of areas, we are aware of
the bias inherent in any such study. 

\subsection{Definitions}
Let $G=(V,E)$ be an undirected multigraph.
We allow multiple edges between two vertices, and thus $E$ is a
multiset. 
The degree $d(u)$ of a vertex $u$ is the number of incident edges to
$u$, taking into account parallel edges. 

Some networks are bipartite, i.e., their vertex set can be partitioned
into two sets such that all edges connect one set with the other.
An example is an interaction network between users and movies, in which
each edge represents an interaction between a user and a movie. 
Most network measures apply to bipartite networks without problem.  Out
of the network measures covered in this article, the only exception is
the clustering coefficient, which is well-defined and always zero for
bipartite networks.  We will therefore not consider the clustering
coefficient for bipartite networks. 

Let $\mathbf A \in \mathbb R^{|V|\times |V|}$ be the symmetric adjacency matrix of
$G$, containing the multiplicity of the edges. 
Thus, $\mathbf A_{uv}$ equals the number of edges connecting $u$ and
$v$. 
We also consider the diagonal degree matrix $\mathbf D \in \mathbb
R^{|V|\times |V|}$ 
defined by $\mathbf D_{uu}= d(u)$.
Finally, the matrix normalized adjacency matrix $\mathbf Z$ is defined as
$\mathbf Z=\mathbf D^{-1/2}\mathbf A \mathbf D^{-1/2}$. 

\section{Measures of Structural Network Diversity}
\label{sec:measures}
We now review the eleven numerical measures of structural network
diversity. 
All eleven measures are summarized in
Table~\ref{tab:definitions}. 
In addition to these eleven measures of diversity, we will also explore the
average degree $d= 2|E| / |V|$ in our experiments, since it has been shown to evolve
monotonically in the literature \cite{b700,b242}.  

For some measures of connectivity, monotonicity proofs exist, which show
that when an 
edge is added to a connected network, the diversity cannot increase, but
can only decrease or remain constant.  These proofs are only valid for the
connected case, i.e., when an edge is added to a connected network.  
The case of an added node is not considered, as it renders the graph disconnected. 
The proofs will be given along with the definitions of the measures. 
As a trivial example, the average degree $d$ is strictly monotonous since
adding an edge increases 
$|E|$ but not $|V|$. 

\begin{table}[t]
  \caption{
    \label{tab:definitions}
    The eleven measures of diversity considered in this article.
    We also show the average degree as reference measure (first line). 
  }
  \makebox[\textwidth]{
    \centering
    \scalebox{0.85}{
      \begin{tabular}{l @{\,\,} l l l c c c c}
        \toprule
        && \textbf{Aspect} & \textbf{Measure} & \textbf{Symbol} & \textbf{Range} & \textbf{Pr.} & \textbf{Mono.} \\
        \midrule
        \ref{sec:measures} &  & --- & Average degree & $d$ & $(0,\infty)$ &  $\nearrow$ & \checkmark \\ 
        \midrule
        \ref{sec:gini} &     				& Pref.\ att.       & Gini coefficient & $G$ & $[0,1]$ & $\nearrow$ & --- \\
        \ref{sec:jain} & Eq.~\ref{eq:jain}     	& Pref.\ att.	& Jain's index & $J$ & $(0,1]$ &  $\searrow$ & --- \\
    \ref{sec:power} & Eq.~\ref{eq:power}		& Pref.\ att.	&   Power-law exponent & $\gamma$ & $(1,\infty)$ & $\searrow$  & --- \\ 
    \ref{sec:entropy} & Eq.~\ref{eq:entropy}	& Pref.\ att.	& Relative edge distribution entropy & $H_{\mathrm{er}}$ & $[0,\infty)$ & $\searrow$  & --- \\
      \midrule
      \ref{sec:diameter} &        & Connect. & 90-percentile effective diameter & $\delta_{0.9}$ & $(0,\infty)$ & $\searrow$  & \checkmark    \\ 
      \ref{sec:wsd} &              & Connect. & Random walk return probability & $\vartheta_r(n)$ & $[1,\infty)$ & $\searrow$ & --- \\
        \ref{sec:controllability} & Eq. \ref{eq:controllability} & Connect. & Relative controllability  & $C_{\mathrm r}$ & $(0,1]$ & $\searrow$ & \checkmark \\
      \ref{sec:alcon} & Eq.~\ref{eq:alcon}                       & Connect. & Algebraic connectivity & $a$ & $[0,\infty)$ & $\nearrow$         & \checkmark \\
        \midrule
        \ref{sec:clusco} & 	& Link pred. &           Clustering coefficient & $c$ & $[0,1]$ &  $\nearrow$  & --- \\
        \ref{sec:rank} & Eq.~\ref{eq:rank}            & Link pred. & Fractional rank & $\mathrm{rank}_{\mathrm F}$ & $[1,\infty)$ & $\searrow$  & --- \\
          \ref{sec:epower} &          			& Link pred. & Eigenvalue power-law exponent & $\alpha$ & $(1,\infty)$ & $\nearrow$ & --- \\
          \bottomrule
      \end{tabular}
        }
      }
      \textbf{Pr.}
      Predicted trend according to the shrinking diversity
      hypothesis. 
      \\
      \textbf{Mono.}
      A monotonicity proof exists, showing that adding an edge to
      a connected network cannot increase the diversity according to the
      given measure. 
\end{table}

\subsection{Preferential Attachment}
\label{sec:richer}
The temporal evolution of equality measures for the degree distribution
can be predicted under the model of preferential attachment. 
Preferential attachment is a general principle of network growth which
states that new edges will connect to a vertex with a probability that
is proportional to the importance of that vertex~\cite{b439}.
The preferential attachment model thus predicts that nodes with high
degree will receive more edges fast, and thus their degree will grow fast,
while small degrees will only grow slowly.  The result is a long-tailed
degree distribution, in which most degrees are small, and few degrees
are large.  To measure the extent of this \emph{long-tailedness} of the
degree distribution, we use four different measures. 

A related but different type of preferential attachment is
preferential attachment on eigenvectors, and is equivalent to the
spectral network evolution hypothesis~\cite{kunegis:spectral-evolution},
which is exploited in link prediction algorithms and is covered in
Section~\ref{sec:linkpre} together with other link prediction methods.   

In an undirected network,
each edge is attached to two nodes.  We can therefore consider each edge
to belong to the two nodes that the edge connects.  Thus, a network can
be viewed as a distribution of 
half-edges over vertices. The number of edges owned by a vertex is
then equal to the number of neighbors of that vertex, i.e., the
degree. The sum of all degrees in the network thus equals twice the
number of edges in the network. 

The interpretation of the equality of degrees as diversity can be
illustrated with a user--movie network. 
When all movies in that network have been seen by the same number of
people, the diversity of the network is high.  If instead a small
number of movies have been seen by many people and most movies have
been seen by only few people, then the network is not diverse.
Thus, network diversity can be measured by inspecting how far the
distribution of edges to nodes is away from an equitable distribution. 

\subsubsection{Gini Coefficient}
\label{sec:gini}
The Gini coefficient is a measure of the inequality of a distribution,
typically used in economics to measure the inequality of the income
distribution in a country. 
As a diversity measure for networks, we apply it to the degree
distribution, as described in~\cite{kunegis:power-law}.  
A measure of inequality can be interpreted to denote the 
opposite of diversity, since a network in which the distribution of edges
is equal can be understood as having more diversity.  

\paragraph{Related Measures}
The Hoover index (or Robin Hood index) is also a measure from economics
equal to the relative amount of total income that must be
redistributed for the distribution to become fully equal~\cite{b702}.  
Another related measure is the balanced inequality ratio
\cite{kunegis:power-law}, which appears 
in statements of the form ``$X$ percent of the population own $(100-X)$
percent of assets.''  For instance, the well-known 80--20 rule states in
its original form that 80\% of land area in Italy was owned by 20\% of
people, leading to a value of 0.2.  
All three measures are ultimately based on the Lorenz curve, and
were found to correlate highly in our experiments, and thus only the Gini coefficient is investigated. 

\subsubsection{Jain's Index}
\label{sec:jain}
Another index of equality that can be applied to the degree distribution
is the index of Jain~\cite{b740}. 
This measure is used in computer networking 
to measure the fairness of
resource allocation. For instance, it is
used to process Internet packets from different sources equally.  
It is defined as 
\begin{align}
  J &= \frac 
  {\left[\sum_{u \in V} d(u)\right]^2}
  {n \cdot \sum_{u \in V} d(u)^2}. 
  \label{eq:jain}
\end{align}
This index is maximally one for a completely equal distribution of
degrees. The theoretical minimum of this index is $1/|V|$ when all edges
attach to a single node. The minimum realizable by a simple network is 
$4(|V|-1) / |V|^2$, i.e., slightly under $4/|V|$. Thus, this
index can be considered to be in range $(0,1]$ for large networks. 

\subsubsection{Power-law Exponent}
\label{sec:power}
In the area of network analysis, the phrase \emph{degree distribution} is
mostly associated with the phrase \emph{power law}.  This is based on
the observation that in many networks, the number of vertices with
degree $d$ is roughly proportional to the power $d^{-\gamma}$, where the exponent
$\gamma > 2$ is a parameter, called the power-law exponent.  
Power-law degree distributions arise for instance in the preferential
attachment model of Barabási and Albert~\cite{b439}, giving a value of
$\gamma=3$ for the basic preferential attachment model. 
%% Other, related
%% models lead to different values of $\gamma$. 

%% \begin{figure}
%%   \centering
%%   \subfigure[Epinions]{
%%     \includegraphics[width=\wTwo]{plot/degree.ay.epinions}
%%   }
%%   \subfigure[CiteSeer]{
%%     \includegraphics[width=\wTwo]{plot/degree.ay.citeseer}
%%     \label{fig:degree:citeseer}
%%   }
%%   \caption[X]{
%%     Degree distributions with fitted power-law exponent
%%     $\gamma$ of the Epinions trust network
%%     and the CiteSeer citation
%%     network. 
%%     (a)~An apparent power law 
%%     (b)~An example of a distribution which appears to be a power 
%%     law starting at a specific degree. 
%%     \label{fig:degree}
%%   }
%% \end{figure}

%% Power-law degree distributions are often a matter of immense
%% discussion (see e.g.~\cite{b462} and its blog
%% entry\footnote{\href{http://cscs.umich.edu/~crshalizi/weblog/491.html}{http://cscs.umich.edu/$\sim$crshalizi/weblog/491.html}}),  
%% with good arguments that in fact many distributions are in practice misidentified as
%% power laws.  For a typical example,
%% see~\cite{b747}, which shows that many distributions arising
%% in biology are in fact not power laws. 
Since not all degree distributions are precise power laws, the power law
exponent is not strictly defined for all networks. 
%% Formally, exact methods exist to verify whether a given
%% degree distribution is a power law at all~\cite{b462}.  
%
%% Figure~\ref{fig:degree} shows examples of degree distributions from
%% actual online network datasets.  As becomes apparent in these plots, not
%% all networks follow a 
%% power-law distribution.  Some only follow a power law beginning at some
%% minimal degree, while others do not follow power laws at all. 
Nonetheless, an estimation of the exponent is often used as
a numerical network measure.  
%% Values of $\gamma$ are reported for many
%% examples in~\cite{b59} and~\cite{b462}. 
In the experiments of this article, we use the method
described in~\cite[Equation (5)]{b408} to estimate the power-law
exponent, defining the power-law exponent $\gamma$ as
\begin{align}
  \gamma &= 1 + n \left(\sum_{u\in V} \ln \frac {d(u)} {d_{\min}} \right)^{-1},
  \label{eq:power}
\end{align}
in which $d_{\min}$ is the minimum degree in the network. 
Note that this method returns values of $\gamma$ in the range
$(1,\infty)$, i.e., the values may be smaller or equal to two. 

A misconception about the exponent $\gamma$ is that the degrees are more
unequal when its value is high. However, the opposite is true:  The
degrees are more unequal when $\gamma$ is
small~\cite[Fig. 6]{kunegis:power-law}. 
Thus, a shrinking diversity implies a shrinking value of $\gamma$. 

\subsubsection{Relative Edge Distribution Entropy}
\label{sec:entropy}
The entropy is a measure used in thermodynamics to
characterize the \emph{disorder} of a physical
system. In information theory, the entropy is a measure of the quantity of
information.  
In a network, we can compute both the entropy of the edge distribution
as well as the entropy of the degree distribution.  

Given a probability distribution $P(x)$ over a finite set $x\in X$, the
entropy $H$ of $P$ is defined as
$H(P) = \sum_{x\in X} -P(x) \ln P(x)$.
The entropy can be interpreted as a measure of the uniformity of a
distribution:  It is zero when $P(x_0)=1$ for some $x_0$, and
reaches its maximal value of $\ln(|X|)$ for the uniform
distribution $P(x)=1/|X|$ for all $x$.  
We apply the entropy to the distribution of edges 
over vertices to define the edge distribution entropy.  In a graph
$G=(V,E)$, the edge distribution entropy is thus
\begin{align*}
  H_{\mathrm e} &= \sum_{u \in V} - \frac{d(u)}{2|E|} \ln \frac{d(u)}{2|E|}.
\end{align*}
The entropy is nonnegative, and 
its maximal possible value is $\ln |V|$, which is
attained when all nodes $u\in V$ have the same degree
$d(u)=2|E|/|V|$. Thus, the edge distribution entropy $H_{\mathrm e}$ is
a measure of equality.
The edge distribution entropy is called the \emph{entropy of degree
  sequence} (EDS) in~\cite{b681}. 

Because the edge distribution entropy has a maximal value of $\ln |V|$,
we may expect it to be highly correlated to the network's size $|V|$
itself.  Therefore, we normalize it by dividing by $\ln |V|$,
resulting in the relative edge distribution entropy
\begin{align}
  H_{\mathrm{er}} &= \frac 1 {\ln |V|} \sum_{u \in V} -
  \frac{d(u)}{2|E|} \ln \frac{d(u)}{2|E|}. 
  \label{eq:entropy}
\end{align}
By construction, $H_{\mathrm{er}}$ varies in the range $[0,1]$, with
zero denoting complete inequality and one denoting complete equality. 
A slightly different definition of the relative edge distribution
entropy is called the \emph{normalized entropy of degree sequence} (NEDS)
in~\cite{b681}. 
The relative edge distribution entropy $H_{\mathrm{er}}$ is thus a
measure of equality, and we expect it to shrink over time. 

\paragraph{Related Measures}
The Theil index is an economic measure of inequality~\cite{b741} often
used to measure income inequality, and related to the entropy by
$T_{\mathrm T} = H_{\mathrm e} / |V|$. 
Another related measure is the Atkinson index, which adds a parameter that can
be used to define the relative importance of small and large
contributions to the inequality measure~\cite{b742}. 

\paragraph{Entropy of Other Distributions}
In addition to the edge distribution, 
the entropy can also be applied to the degree distribution~\cite{b693}. 
This entropy is invariant under exchanges of 
the number of nodes having any different degree values $d_1$ and
$d_2$, and thus two very different degree distributions could share the same entropy value.  Thus, it should only be used under specific circumstances, such as the network having a power-law degree
distribution, a problem which it shares with the power-law exponent
$\gamma$. 

\subsection{Connectivity}
\label{sec:connect}
The concept of connectivity characterizes a network whose nodes are easily reachable
from other nodes.  Different definitions of \emph{easily reachable} lead
to different measures of connectivity. For instance, counting the
maximal number of edges needed leads to the diameter, and measuring how
likely a random walk of $n$ steps returns to its starting node leads to
the random walk return probability. 
A measure of connectivity can be interpreted as a measure of diversity
in the following way:
When connectivity is high, any part of the network is easily
reachable from any other part, and thus the network lacks diversity.
On the other hand, a network with a low connectivity value has more
subgraphs well-separated from the rest of the network, and thus more
local diversity. 

We study four measures of connectivity:  the diameter, which corresponds
to the maximal length of shortest paths in the network; the random walk
return probability, which corresponds to the probability of a random walk to
return to its starting node; the relative controllability, based on the
number of nodes needed to control a full network; and the
algebraic connectivity, based on a spectral clustering of the network. 

\subsubsection{Diameter}
\label{sec:diameter}
The diameter is a very common network measure that equals the longest
shortest path in the network. 
It is typically used to describe a network as a \emph{small
  world}~\cite{b228}.  A small-world network is one in which the diameter is
small, and the clustering coefficient is high.  The intuition behind
pairing these two measures is to combine a measure of local coherence,
the clustering coefficient, with a measure of the overall coherence, the
diameter.  The given reference shows that the diameter places each
network on a continuum between two extremes. 
On one hand, a lattice graph has a high local
coherence, and thus a high clustering coefficient, but a low global
coherence, and thus a large diameter.  It could be said that the lattice
has a high diversity, since its parts are very far from each other, as
measured by the typical distance of nodes.  On the other hand, a 
random graph has a low clustering coefficient and a low diameter, denoting low
diversity, due to the fact that every node is reachable in few hops from
every node.  This is consistent with the interpretation of the random
graph as having low diversity, since all nodes are near to each other,
and thus any local structure is lost. Therefore, the diameter of a
network can be considered a measure of the network diversity. 

To be precise, the diameter measures the largest distance between two nodes of a
network.  
In
practice, the diameter is susceptible to long branches connected to the
rest of the network on just one end, and therefore a common variant is the
90-percentile effective diameter, defined as the number of steps needed
to reach 90\% of all nodes, counted over all nodes.
We refer to this graph property as $\delta_{0.9}$, and will use it in
the experiments of this article. 

One important restriction of the diameter is that it can only be applied
to a connected network.  For an unconnected network, some node pairs are
not reachable from each other, and the diameter is undefined or
infinite.  Therefore, we always measure the diameter on a network's
largest connected component.  
In \cite{b771} and \cite{b242}, the diameter is observed to shrink over
time, implying 
that the diversity of the network is becoming less over time.
Due to the high runtime complexity of computing the exact effective
diameter, we estimate it by sampling vertices, and computing their
distance to all other vertices.

\paragraph{Monotonicity}
Adding an edge to a connected network cannot increase the distance
between nodes, and thus the diameter can only decrease or remain
constant when such an edge is added. 
The diameter is thus monotonous with regard to adding an edge to a
connected network.  Note that this does not hold for unconnected
networks.  In the general case, adding an edge to an unconnected network
may increase the diameter of the largest connected component. 

%% \paragraph{Related Measures}
%% A related network measure is the radius, i.e., the shortest
%% eccentricity of all nodes, where the eccentricity of a node is defined
%% as the longest distance from that node to any other node.  Like the
%% diameter, the radius is skewed by the presence of long branches. 

\subsubsection{Random Walk Return Probability}
\label{sec:wsd}
Another way to measure connectivity in a network is to consider random
walks.  A random walk is a process starting at a given node $u$ and
proceeding along edges in a random manner.  At each node $v$, the random walk
chooses one of the neighbors of that node uniformly at random;
i.e., with probability equal to $1/d(v)$.  Random walks 
can be used to measure how well connected a network is.  For instance,
one can consider the probability of return to the starting node $u$
after $n$ steps.  If it is low, then the network possesses low locality,
which we interpret as a high connectivity and thus low diversity.  If
the probability of returning to 
the initial node is high, then we interpret that as a sign of
low connectivity and thus high diversity.

The random walk return probability was introduced by Fay et al.\ \cite{b425} 
as a metric for comparing two graphs defined by
\begin{align*}
  \vartheta(n) & =
  \sum_k(1-\lambda_k[\mathbf Z])^n=\sum_C \frac 1 {d(u_1)d(u_2)\ldots d(u_n)},
\end{align*}
where
$C$ is the set of all 
cycles of size $n$ in the graph\footnote{The authors in~\cite{b425}
  recommend a value of $n= 4$, 
  which we use in this article.} and $d(u_i)$ is the degree of the
$i^{\mathrm{th}}$ node in a cycle.

To compute the random walk return probability $\vartheta (n)$, we thus need
to compute all eigenvalues of the matrix $\mathbf Z$, or alternatively to enumerate all $n$-cycles.
Both operations are expensive, and cannot be achieved in practice
on large datasets.  Therefore, we use, as a proxy, 
the sum $\vartheta_r$ only over the $r$ dominant
eigenvalues,
i.e., those with
the greatest distance to $\lambda=1$. 
This is not an approximation to $\vartheta(n)$, but a value
that varies in conjunction with it, both reproducing
shifts in the overall distribution of eigenvalues in the range
$[0,2]$.  

As a network evolves, a shrinking random walk return probability shows that
this sum is shrinking and so the probability of taking a random walk of
length $n$ and returning to the source node is in general shrinking
in the network. Another way of expressing this is that the number of
\emph{escape routes or non-cycles} has increased. This in turn occurs
when the community structure of networks becomes more blurred; random
walks are more likely to jump away from the community where they
started. Thus the lower the random walk return probability, the lower the diversity of a
network.

\subsubsection{Controllability} 
\label{sec:controllability}
A less-known way to assess the structure of a network consists in
measuring how well it can be controlled.  For instance, assume that we
want to influence opinions in a social network, but are only able to
directly influence $k$ persons in the network, much less than the number
of vertices $|V|$. Assuming that opinions
will spread through the network, how big has $k$ to be in order for us
to be able to influence all nodes in a network, in a way that any
arbitrary opinion can be given to any node?  A solution to this problem
is given by Liu et al.\ \cite{b673}, in which such driver nodes are
identified and, surprisingly, they are not necessarily the nodes with
highest degree.  In fact, the authors of that article state that driver
nodes tend to \emph{avoid the hubs} of the network. 

The resulting computational model uses differential equations to model
diffusion and can be reduced to finding a maximal matching in the
bipartite double cover of the network~\cite{b673}. The maximal matching
in a bipartite graph can be computed efficiently by exploiting König's
theorem, which states that perfect matchings and minimal vertex covers have
equal size in bipartite graphs~\cite{b748}, and thus the corresponding integer
program formulations are equivalent to their relaxations, implying that
the two problems can be solved in polynomial time.  
In fact, a maximal matching in a bipartite graph can be found in runtime
$O(|V|^{1/2}|E|)$, and thus can be computed efficiently even
for large networks. 

The number of driver nodes $C$ needed to control a graph $G=(V,E)$
equals $|V|$ minus the size of the 
maximal directed 2-matching in the network.
A 2-matching is a set
of edges such that each vertex is incident to at most two edges.  A
directed 2-matching is a set of directed edges, such that each vertex is
incident to at most one ingoing and one outgoing edge.  Here, we
interpret an undirected graph as a directed graph where each edge
corresponds to two directed edges:
\begin{align*}
  |V| - C &= \max_{M \subseteq V^2} |M| \\
  \text{s.t.} & \quad \left| \left\{ (v,w) \in M \mid v = u, \{v,w\}\in
  E\right\}
  \right| \leq 1 
  & \text{for all } u \in V, \\
  &  \quad \left| \left\{ (v,w) \in M \mid w = u, \{v,w\}\in E  \right\}
  \right| \leq 1 
  & \text{for all } u \in V. 
\end{align*}
The result is the number $C$ of vertices needed to control a given
network.  A network that is hard to control (i.e.,
has a high value $C$) can be interpreted as having a low connectivity
and thus a higher diversity.  Thus, we expect
$C$ to be a measure of the diversity of a network. 

Since the number of nodes in a network is changing over time, the number
of driver nodes and thus $C$ is dependent on the size of the
network. Therefore, we use as a measure of diversity the relative
controllability $C_{\mathrm r}$ defined as
\begin{align}
  C_{\mathrm r} &= \frac C {|V|}. 
  \label{eq:controllability}
\end{align}

\paragraph{Monotonicity}
When an edge is added to a connected graph $G=(V,E)$, the size of a
maximal matching in its bipartite double cover can only increase or
remain constant, and therefore the number of driver nodes $C$ and the
relative 
controllability $C_{\mathrm r}$ can only
decrease or stay constant, but not increase. 

\subsubsection{Algebraic Connectivity}
\label{sec:alcon}
In some graphs, removing a single edge can make the graph disconnected.
These kinds of graphs have low connectivity.  On the other hand, some
graphs can only be made disconnected by removing a much larger number of
edges.  These kinds of graphs have a high connectivity.
To measure these differences, 
the edge connectivity of a graph can be defined as the number of edges
that have to be removed to make the graph disconnected~\cite{b116}.  
This number is
a characteristic number of the graph.  However, it is not very
expressive.  For instance, the edge connectivity can be reduced to a 
value of one by
adding to the graph a new vertex and a new edge between that vertex and an already
existing vertex.
Instead, a more robust measure of connectivity is the algebraic
connectivity, which 
is based on 
the Laplacian matrix 
$\mathbf L=\mathbf D-\mathbf A$.
The matrix $\mathbf L$ is positive-semidefinite.  All its eigenvalues
are nonnegative, and its smallest eigenvalue is zero. Its second
smallest eigenvalue is a measure of the network's connectivity:  When
the network is disconnected, it is zero too.  Otherwise, it is larger
the harder it is to find small cuts dividing the network into two
parts. We will denote the algebraic connectivity
\begin{align}
  a = \lambda_2[\mathbf L],
  \label{eq:alcon}
\end{align}
where $\lambda_2[\mathbf L]$ is the second-smallest eigenvalue of
$\mathbf L$. 
The algebraic connectivity was initially defined by Fiedler
\cite{b652}.  
We compute the algebraic connectivity only in the largest connected
component of a graph, as we do for the diameter. Otherwise, only the
algebraic connectivity of the smallest 
connected component would be considered, since the eigenvalues of
$\mathbf L$ are the eigenvalues of the Laplacian of each connected
component. 

\paragraph{Monotonicity}
%% The evolution of the algebraic connectivity 
%% can be explained by considering newly created vertices.
%% When a vertex is added to a network, it is not connected to any node in
%% the network.  Therefore, the matrix
%% $\mathbf L$ only changes by addition of a zero row and a zero column.
%% This operation makes the graph unconnected and therefore the algebraic
%% connectivity becomes zero, although the algebraic connectivity of the
%% largest connected component does not
%% change. 
%% When however a first edge is added to a new vertex, we can
%% show that the algebraic connectivity can only decrease.  
%% On the other
%% hand, 
Adding an edge between two vertices that are already connected indirectly can
only increase the algebraic connectivity of a graph.
This result can be deduced by considering the following interlacing
theorem, stated e.g.\ in~\cite[p.~97]{b663}.  This
theorem states that when adding a
positive-semidefinite matrix $\mathbf E$ of rank one to a given
symmetric matrix $\mathbf X$
with eigenvalues $\lambda_1 \leq \lambda_2 \leq \ldots \leq
\lambda_n$, the new matrix $\mathbf{\tilde X} = \mathbf X + \mathbf E$ has
eigenvalues $\tilde \lambda_1 \leq \tilde \lambda_2 \leq \ldots \leq
\tilde \lambda_n$ which interlace the eigenvalues of $\mathbf X$:
\begin{align*}
  \lambda_1 \leq \tilde \lambda_1 \leq \lambda_2 \leq \tilde\lambda_2 \leq \ldots
  \leq \lambda_n \leq \tilde\lambda_n
\end{align*}
In the case of the Laplacian matrix $\mathbf L$, adding an edge
$\{u,v\}$ adds to it the rank-one matrix $\mathbf x \mathbf x^{\mathrm T}$,
where $\mathbf x\in \mathbb R^{|V|}$ is a vertex vector defined by
$\mathbf x_{u}=+1$, $\mathbf x_{v}=-1$ and $\mathbf x_w=0$ otherwise.
Note that the chosen orientation of the edge does not matter in this
definition.  Thus, a positive-definite matrix is added to $\mathbf L$,
and the spectrum of $\mathbf L$ thus shifts up. 

If $G$ is already connected, its spectrum is $\{0,
\lambda_2, \ldots\}$ and after addition of the edge $\{u,v\}$ it becomes
$\{0, \tilde \lambda_2, \ldots\}$, from which it follows that $\tilde
\lambda_2 \geq \lambda_2$, i.e., the algebraic connectivity can only
grow or remain constant, but not decrease.

\subsection{Link Prediction Functions}
\label{sec:linkpre}
The problem of link prediction in networks consists in predicting which
edges will appear in an evolving network, given the current
network. Many link prediction functions exist, each corresponding to
different models of network growth. Link prediction functions make
assumptions of regularity in a network. For instance, the 
triangle closing model is based on the assumption that triangles will
form in a network. Thus, triangle closing predicts a
growing regularity in a network. This is also true for other link
prediction functions, as they tend to predict common structures, and
therefore predict that common structures will be reinforced while
uncommon structures will stay uncommon. 
Regularity can be interpreted as an 
aspect of diversity, in the sense that regularity indicates the lack
of structural diversity. Thus, link prediction functions can be
interpreted as predicting an increasing regularity and a shrinking
diversity. 

We consider three link prediction functions which lead to three measures of
diversity:  the clustering coefficient, the fractional rank and the
eigenvalue power-law exponent. 

\subsubsection{Clustering Coefficient}
\label{sec:clusco}
The clustering coefficient measures the fraction of adjacent edge pairs
that are completed by a third edge to form a triangle.  The tendency of
networks to form triangles represents one half of the small-world
network model along with the network diameter~\cite{b228}, and leads to
the simplest network growth 
models that goes beyond preferential attachment to take into account the
shared neighborhood of two nodes~\cite{b461}:  triangle closing, i.e.,
the prediction that new edges will appear such that many triangles are
formed.  As a link 
prediction function, the resulting \emph{common neighbor count} function
is one of the simplest 
possible link prediction methods.

The clustering coefficient is the only measure we consider that only makes sense for
unipartite networks.  For bipartite networks, it is zero, because a
bipartite network does not contain triangles. 

In the triangle closing model, new edges are predicted to form new
triangles and thus, the clustering coefficient is expected to increase,
and thus the diversity to shrink. 

\subsubsection{Fractional Rank}
\label{sec:rank}
An important class of link prediction functions are graph
kernels.  Graph kernels are positive-semidefinite functions of the
adjacency matrix $\mathbf A$ 
and can be used for modeling network growth~\cite{b137}. 
%% An example is the exponential kernel $\exp(\alpha \mathbf A)$~\cite{b156}. 
Graph kernels as considered here have the property that they can be
expressed in terms of 
the eigenvalue decomposition of the matrix $\mathbf A$. 
Let 
$\mathbf A = \mathbf U \mathbf \Lambda \mathbf U^{\mathrm T}$
be the eigenvalue decomposition of $\mathbf A$, in which $\mathbf U$ is
an orthogonal matrix and $\mathbf \Lambda$ a diagonal matrix. Then, a
graph kernel $F$ 
can be expressed as a function of the form
$F(\mathbf A) = \mathbf U F(\mathbf \Lambda) \mathbf U^{\mathrm T}$,
where $F(\mathbf \Lambda)$ is given by applying a function $f$ to each
eigenvalue separately such that $(F(\mathbf \Lambda))_{kk} = f(\mathbf
\Lambda_{kk})$ \cite{kunegis:spectral-transformation}. 
Common graph kernels of this form have the property that the function
$f$ is convex, i.e., they make large eigenvalues grow faster than
smaller ones. Thus, large eigenvalues will tend to dominate smaller ones
if a network evolves according to such graph kernels. 

The two main graph kernels we study are the exponential
kernel~\cite{b156}
$e^{\alpha \mathbf A} = \mathbf U e^{\alpha \mathbf \Lambda} \mathbf U^{\mathrm T}$
and the Neumann kernel~\cite{b263}
$(\mathbf I - \alpha \mathbf A)^{-1} = \mathbf U (\mathbf I - \alpha
\mathbf \Lambda)^{-1} \mathbf U^{\mathrm T}$. 
Both kernels take a positive parameter $\alpha$. For the Neumann kernel,
this parameter must be smaller than the inverse spectral norm of
$\mathbf A$, i.e., smaller than the inverse of the largest absolute
eigenvalue of $\mathbf A$. 

Let $|\lambda_1| \geq |\lambda_2| \geq |\lambda_3| \geq \dotsb$ be the
ordered eigenvalues of $\mathbf A$, i.e., the diagonal elements of
$\mathbf \Lambda$.  From the convexity of the function $f$, it follows
that the ratio $|\lambda_k|/|\lambda_1|$ and its square shrink during
the application of a graph
kernel, and therefore both the absolute and the squared eigenvalue
sums are predicted to shrink if graph kernels are correct link
prediction functions.  

The sum of absolute or squared eigenvalues of the adjacency matrix can
be derived as a fractional extension of the rank
of a matrix. These measures are nonnegative and generalize the notion
of matrix rank as follows. 
The ordinary matrix rank of $\mathbf A$ can be written as
$\mathrm{rank}(\mathbf A) = \sum_k [\lambda_k \neq 0]$,
in which $[\lambda \neq 0]=1$ when $\lambda\neq 0$ and $[\lambda \neq 0]=0$
otherwise. 
We thus see that the matrix rank counts 
the number of nonzero eigenvalues.  This is clearly a measure of the
diversity of the network, but not a very good one, because very small
eigenvalues contribute a value of one, although their
contribution to the network is very small.

Therefore, we propose to compute
a \emph{fractional rank} in which each eigenvalue is counted in
proportion to its size~\cite{kunegis:network-rank}.  We start with the largest eigenvalue
$\lambda_1$ and define its weight to be one.  Then, each subsequent
eigenvalue $\lambda_k$ is weighted as $(\lambda_k / \lambda_1)^2$.  The
sum of these values then gives the network rank:
\begin{align}
  \mathrm{rank}_{\mathrm F} &= \sum_k \left(\frac {\lambda_k}
         {\lambda_1}\right)^2 
         \label{eq:rank}
\end{align}
We can rewrite this as the ratio of the Frobenius norm $\|\mathbf
A\|_{\mathrm F}$ and the spectral norm $\|\mathbf A\|_2$ of $\mathbf
A$:
\begin{align*}
  \mathrm{rank}_{\mathrm F} 
  = \left(\sum_k \lambda_k^2\right) / \lambda_1^2
  = \frac {\left\| \mathbf A \right\|_{\mathrm F}^2} {\left\| \mathbf A
    \right\|_2^2} 
  = \frac {2 |E|} {\lambda_1^2}
\end{align*}
This is true because the spectral norm equals the largest
absolute eigenvalue $|\lambda_1|$, and the Frobenius norm equals the
square root the the sum of squared eigenvalues of $\mathbf A$.\footnote{
Note that $\lambda_1 \geq 0$ in our case, because the entries of $\mathbf A$ are
nonnegative.   }
We will call this number the fractional rank of $G$.
Note that because the squared Frobenius norm $\left\|\mathbf
A\right\|_{\mathrm F}^2$ equals the sum of squared
eigenvalues, we have $\mathrm{rank}_{\mathrm F}(\mathbf A) \geq 1$. 
The fractional rank can be easily computed using the
number of edges in the graph and the spectral norm, because $\|\mathbf A
\|_{\mathrm F}^2 = \sum_{i,j} \mathbf A_{ij}^2 = 2|E|$.
The spectral norm $\|\mathbf A\|_2$ equals the largest absolute value 
and can be computed by power iteration. 

\paragraph{Preferential Attachment on Eigenvectors}
A shrinking fractional rank can also be explained by a modification of the
preferential attachment model:  The eigenvector centrality preferential
attachment model, which states that the probability that an edge attaches to a
vertex is proportional to that node's eigenvector centrality.  The
eigenvector centrality is a centrality measure for nodes in a network,
based on the eigenvalue decomposition of the network's adjacency
matrix. It is defined as the vertex's entry in the adjacency matrix's
dominant eigenvector.  This value is always nonnegative as a result of the
Perron--Frobenius theorem. 

When an unconnected node is added to a network, the fractional rank does not
change.  This follows directly from the fact that adding a zero row and
column to a matrix will add an eigenvalue of zero to the spectrum.
When an edge is added, the situation is more complex. In the case of the
fractional rank $\mathrm{rank}_{\mathrm F}$ of a graph $G=(V,E)$ we can make the following
derivation.  
Let $\tilde G=(V, E \mathbin{\cup} {\{u,v\}})$ be the graph $G$ to which the edge
$\{u,v\}$ has been added.
Also, let $\mathbf{\tilde A}$ be
its adjacency matrix.
Then, the new largest eigenvalue $\tilde \lambda_1$ can be estimated in
the following way~\cite{b508}.  Let $\mathbf e_u\in \mathbb R^{|V|}$ be
the vertex vector defined by $(\mathbf e_u)_v=1$ when $u=v$ and
$(\mathbf e_u)_v=0$ otherwise.
Also, let $\mathbf A=\mathbf U \mathbf \Lambda \mathbf U^{\mathrm T}$ be
the eigenvalue decomposition of $\mathbf A$.
Then, the new adjacency matrix $\mathbf{\tilde
  A}$ can be written as
$\mathbf{\tilde A} = \mathbf A + \mathbf e_u^{\phantom{\mathrm I}} \mathbf e_v^{\mathrm T} +
  \mathbf e_v^{\phantom{\mathbf I}} \mathbf e_u^{\mathrm T}$.
Now, assuming we want to write the new adjacency matrix as
$\mathbf{\tilde A}$ as $\mathbf{\tilde A}=\mathbf U \mathbf{\tilde
  \Lambda}\mathbf U^{\mathrm T}$, we get
$\mathbf{\tilde \Lambda} = \mathbf \Lambda +
  \mathbf U^{\mathrm T} ( \mathbf e_u^{\phantom{\mathrm I}} \mathbf e_v^{\mathrm T} + \mathbf
  e_v^{\phantom{\mathrm I}} \mathbf e_u^{\mathrm T}) \mathbf U$.
The matrix $\mathbf{\tilde \Lambda}$ defined in this way is not
diagonal. However, in practice it is usually almost diagonal under the
spectral network evolution model~\cite{kunegis:spectral-evolution}, and its largest
diagonal value 
can be estimated as
\begin{align*}
  \tilde \lambda_1 &= \lambda_1 + \mathbf U_{u1} \mathbf U_{v1} +
  \mathbf U_{v1} \mathbf U_{u1}
\end{align*}
The meaning of this expression is that approximately, by adding the edge
$\{u,v\}$, the dominant
eigenvalue of the adjacency matrix $\mathbf A$ will grow by the double
of the product of the entries $u$ and $v$ of the dominant eigenvector of
$\mathbf A$.  Plugging this result into the definition of the fractional
rank, it follows that $\mathrm{rank}_{\mathrm F}$ shrinks 
when
\begin{align*}
  \frac {2|E|}{\lambda_1} > \frac {2(|E|+1)}{\lambda_1 + 2 \mathbf
    U_{u1}\mathbf U_{v1}},
\end{align*}
or equivalently when
$\mathbf U_{u1} \mathbf U_{v1} > \lambda_1 / 2|E|$.
In other words, the fractional rank shrinks when the values
$\mathbf U_{u1}$ and $\mathbf U_{v1}$ are large enough.  Remember that
the dominant eigenvector $\mathbf U_{\bullet 1}$ of $\mathbf A$ is
nonnegative and can be
interpreted as the eigenvector centrality of nodes in $G$.
Thus, the fractional rank shrinks when the product of the
eigenvector centralities of the connected vertices are large enough.
This can be understood as a form of preferential attachment:  When new
edges connect to central nodes, the fractional rank shrinks.  The
difference with the preferential attachment model is in the choice of
the eigenvector centrality instead of the degree centrality.

\paragraph{Spectral Growth}
Another way to analyse the evolution of the fractional network rank is to
look at models predicting the evolution of the largest eigenvalue
$\lambda_1$.  It follows from the definition of $\mathrm{rank}_{\mathrm F}$ that
the fractional network rank shrinks when the largest eigenvalue $\lambda_1$
grows faster than the square root of the number of
edges $|E|$. A corresponding model is given in~\cite{b537}, where the
largest eigenvalue $\lambda_1$ grows as $|V|^{1/4}$.  According
to~\cite{b242}, the number of edges $|E|$ grows super-linearly in the number
of vertices $|V|$, i.e., there is a constant $c>1$ such that
$|E|\sim|V|^c$.  Plugging this into the definition of the fractional
network rank, we get
\begin{align*}
  \mathrm{rank}_{\mathrm F}
  &= \frac {2|E|}{\lambda_1^2}
  \sim \frac {|V|^c} {\lambda_1^2}
  \sim \frac {|V|^c} {(|V|^{1/4})^2}
  = |V|^{c-1/2}.
\end{align*}
Thus, the fractional network rank will shrink when $c < 3/2$.
Coincidentally, the constant $c$ has been reported to vary between $1.1$
and $1.7$.  
This is consistent with our experiments, in which the fractional rank does not
\emph{always} shrink, but only in most cases. 

\paragraph{Linear Spectral Evolution}
The spectral evolution model from~\cite{kunegis:spectral-network-evolution}
implies that the fractional rank decreases. 
In this model, it is assumed that over time, only the eigenvalues of the
adjacency matrix $\mathbf A$ grow, and that the eigenvectors
of $\mathbf A$ stay constant.  Specifically, it predicts that the
evolution of each eigenvalue is linear.  

If spectral growth is extrapolated linearly into the future, the
eigenvalue with the largest growth rate will overtake all others, and
the network rank will decrease until is reaches the number of
eigenvalues that have the same maximal growth rate. This explains a
shrinking fractional rank in many networks, as a single eigenvalue
becomes dominant. 

\subsubsection{Eigenvalue Power-law Exponent}
\label{sec:epower}
\begin{figure}[t]
  \centering
  \subfigure[Enron]{
    \includegraphics[width=\wTwoPointFive]{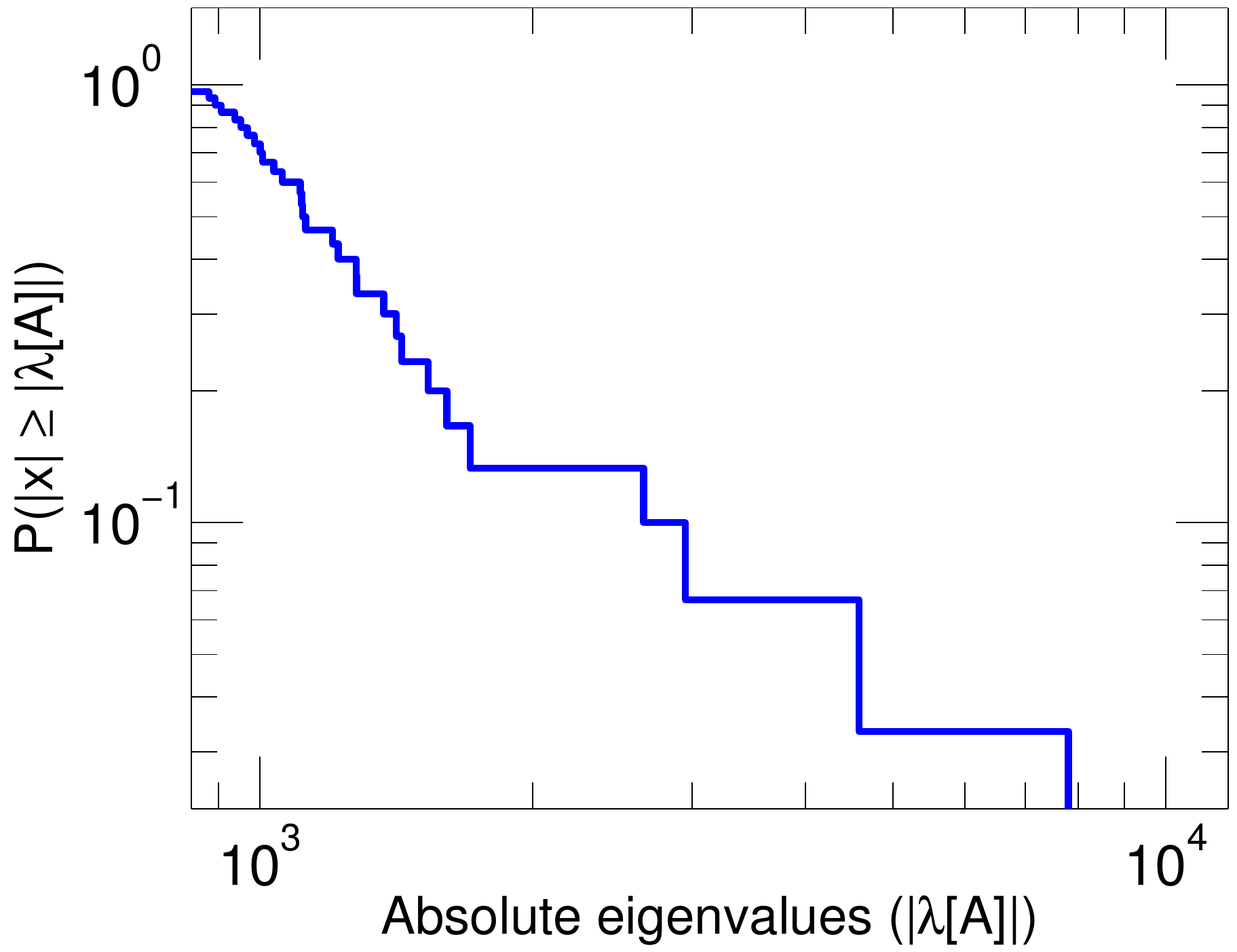}}
  \subfigure[Epinions trust]{
    \includegraphics[width=\wTwoPointFive]{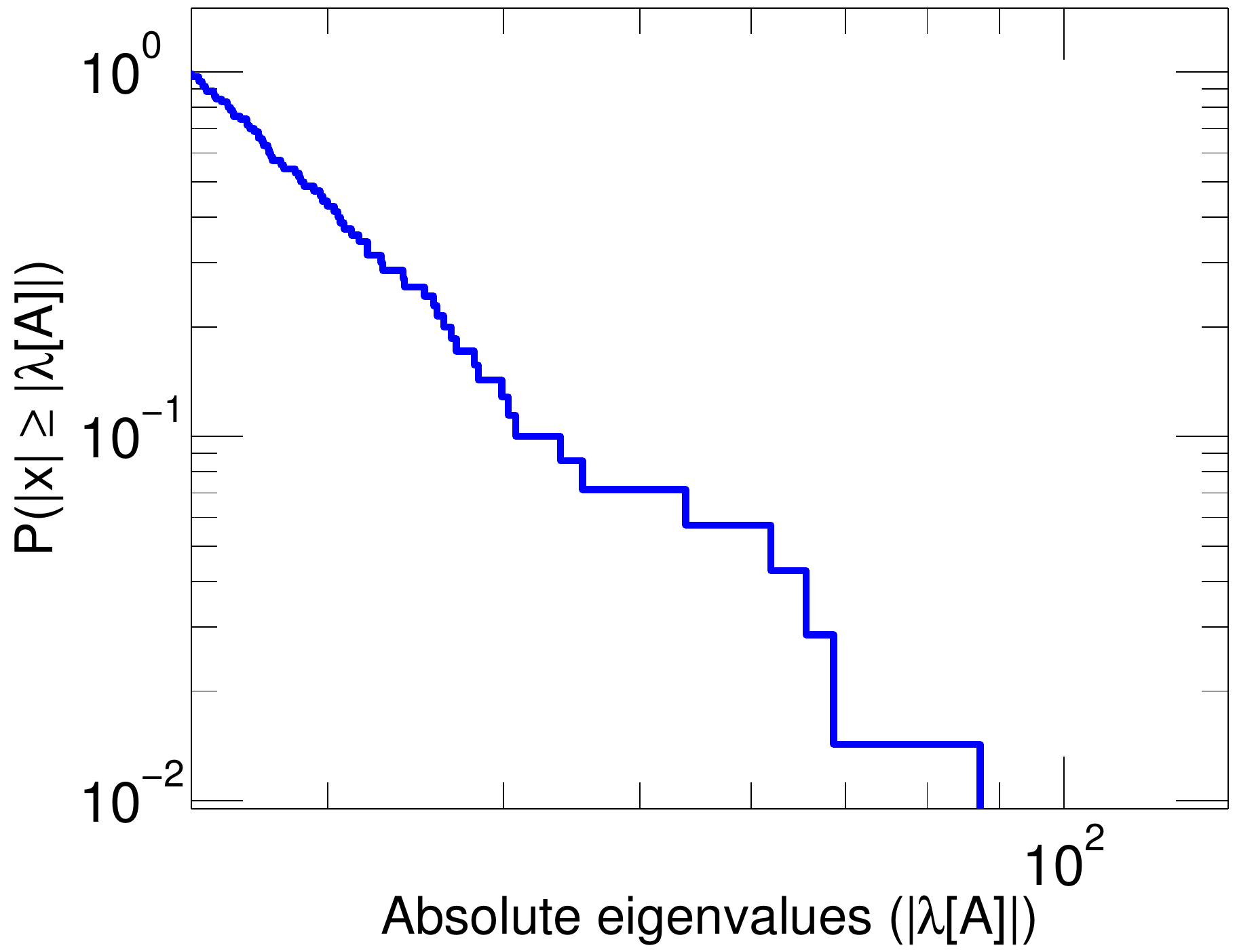}}
  \caption{
    Examples of eigenvalue power laws. The plots show the cumulated
    distribution of the eigenvalues of the adjacency matrix $\mathbf A$,
    on a log-log scale. 
  }
  \label{fig:epower-example}
\end{figure}

Another way to measure the effect of graph kernels on growth of networks
is given by the eigenvalue power-law exponent. The largest eigenvalues
of a network's adjacency matrix almost always follow a long tailed
distribution. In other words, there are much more small eigenvalues than
large eigenvalues. 
In~\cite{b535}, the distribution of the
largest eigenvalues of adjacency matrix $\mathbf A$ of the Internet
topology network where observed to follow a
power law $\lambda_k[\mathbf A] \approx \lambda_1[\mathbf A] / \alpha^k$
with $\alpha > 1$.
Examples are shown in Figure~\ref{fig:epower-example}. 

\begin{figure}[t]
  \centering
  \includegraphics[width=\wTwo]{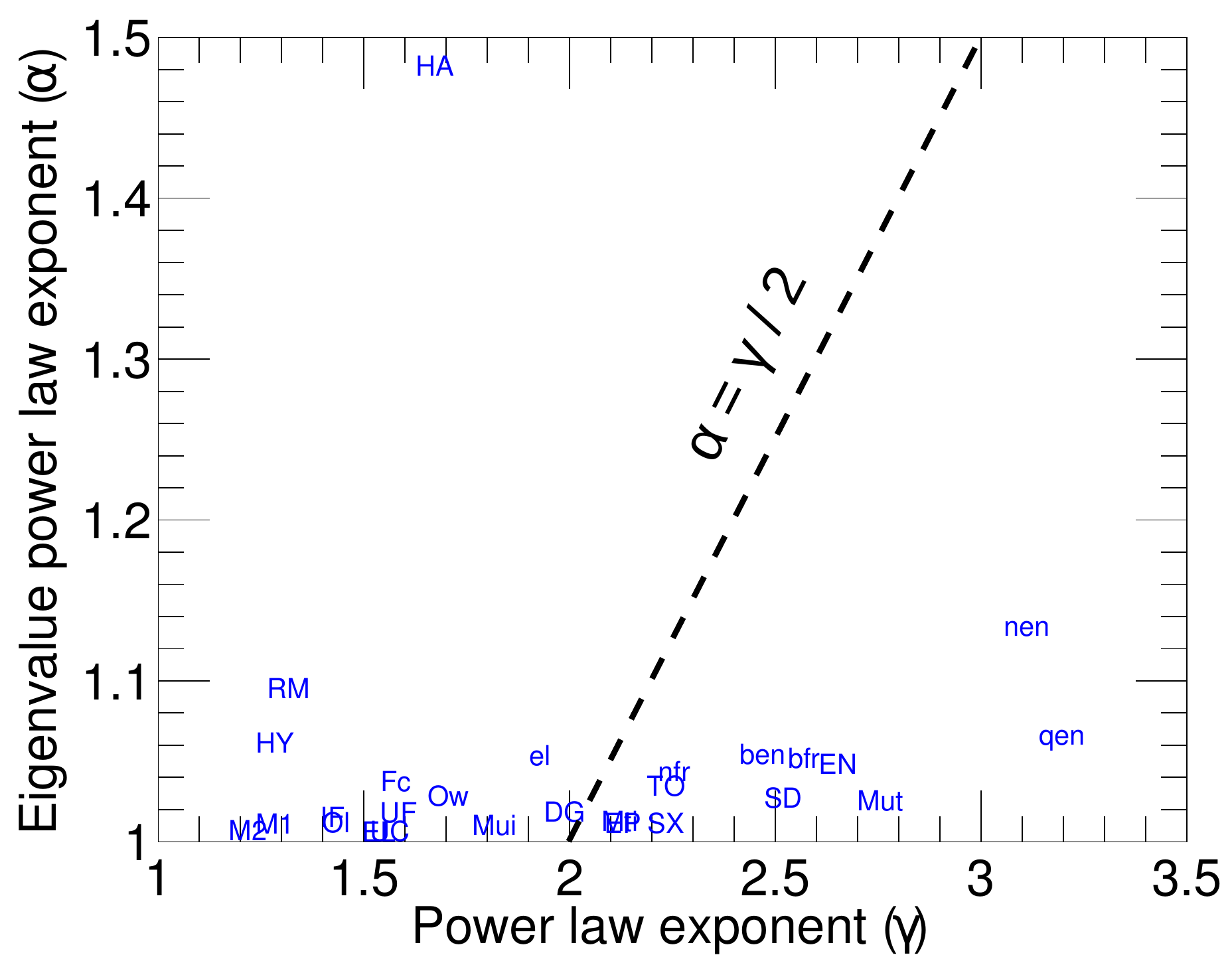}
  \caption[*]{
    The degree power-law exponent $\gamma$ compared with the eigenvalue
    power-law exponent $\alpha$ for all datasets.
    This plot shows no correlation between the two measures, in
    contradiction to results in~\cite{b672} (shown as a dashed line).
  }
  \label{fig:scatter.power.epower}
\end{figure}

As shown in the previous section, graph kernels predict that large
eigenvalues grow faster than smaller eigenvalues. Thus, the exponent $\alpha$ of
eigenvalue power laws is expected to grow and
we expect the value $\alpha$ to grow when the diversity of a
network is shrinking. 
We estimate the power-law exponent $\alpha$ by using the method 
described in~\cite[Equation (5)]{b408}. This is the same method we use
to estimate the degree power-law exponent $\gamma$ as described in
Section~\ref{sec:power}.  
The degree power-law exponent and the eigenvalue power-law exponent are
derived to be related in~\cite{b672} by the expression $\alpha = \gamma
/ 2$.
This is
explained in~\cite{b320} by the observation that the largest eigenvalues of
a scale-free graph follow the square roots of the largest degrees.
Figure~\ref{fig:scatter.power.epower} shows a scatter plot of both
exponents for all datasets. 
This plots shows no correlation between the two power-law exponents.
Also, the fact that $\gamma$ is a measure of diversity while $\alpha$ is
a measure of non-diversity 
is consistent with no such linear relationship. 

\section{Experiments}
\label{sec:experiments}
To validate the hypothesis of shrinking structural diversity, we use the
Koblenz Network Collection
(KONECT\footnote{http://konect.cc/~\cite{konect}}),  
which, at the time of the experiments, consisted of 187 network 
datasets, of which  
72 had information about edge creation times. Out of these datasets, we
use twenty-seven datasets for which we were able to compute all measures in
reasonable time. Of these, thirteen are unipartite and fourteen bipartite. 
For directed and
weighted networks, we ignore edge directions and edge weights. 
In all networks used, the edges are labeled by edge creation times. 
The full list of network datasets used is given in
Table~\ref{tab:datasets}. 

\begin{table}[t]
  \caption{
    \label{tab:datasets}
    The list of twenty-seven network datasets used in this study.
  }
  \centering
  \scalebox{0.7}{
    \begin{tabular}{r l l l r r }
      \toprule
      &   & \textbf{Network} & \textbf{Flags} &
      \textbf{$|V|$} & \textbf{$|E|$} \\
      \midrule
      \cite{download.wikimedia.org} & \textsf{ben} & Wikibooks, English & B & 167,525 & 1,164,576 \\
\cite{download.wikimedia.org} & \textsf{bfr} & Wikibooks, French & B & 30,997 & 201,727 \\
\cite{b565} & \textsf{DG} & Digg & U M & 30,398 & 87,627 \\
\cite{download.wikimedia.org} & \textsf{el} & Wikipedia, Greek & B & 149,904 & 1,837,141 \\
\cite{b551} & \textsf{EL} & Wikipedia elections & U M & 8,297 & 107,071 \\
\cite{b345} & \textsf{EN} & Enron & U M & 87,273 & 1,148,072 \\
\cite{b367} & \textsf{EP} & Epinions trust & U M & 131,828 & 841,372 \\
\cite{said:social-similarity} & \textsf{Fc} & Filmtipset & B & 75,360 & 1,266,753 \\
\cite{b532} & \textsf{HA} & Haggle & U & 274 & 28,244 \\
\cite{konect:sociopatterns} & \textsf{HY} & Hypertext 2009 & U & 113 & 20,818 \\
\cite{konect:sociopatterns} & \textsf{IF} & Infectious & U & 410 & 17,298 \\
\cite{www.grouplens.org/node/73} & \textsf{M1} & MovieLens 100k & B & 2,625 & 100,000 \\
\cite{www.grouplens.org/node/73} & \textsf{M2} & MovieLens 1M & B & 9,746 & 1,000,209 \\
\cite{www.grouplens.org/node/73} & \textsf{Mti} & MovieLens tag–movie & B & 24,129 & 95,580 \\
\cite{www.grouplens.org/node/73} & \textsf{Mui} & MovieLens user–movie & B & 11,610 & 95,580 \\
\cite{www.grouplens.org/node/73} & \textsf{Mut} & MovieLens user–tag & B & 20,537 & 95,580 \\
\cite{download.wikimedia.org} & \textsf{nen} & Wikinews, English & B & 173,772 & 901,416 \\
\cite{download.wikimedia.org} & \textsf{nfr} & Wikinews, French & B & 26,546 & 193,618 \\
\cite{b480} & \textsf{Ol} & Facebook friendships & U & 63,731 & 1,545,686 \\
\cite{b480} & \textsf{Ow} & Facebook wall posts & U M & 63,891 & 876,993 \\
\cite{download.wikimedia.org} & \textsf{qen} & Wikiquote, English & B & 116,363 & 549,210 \\
\cite{b665} & \textsf{RM} & Reality Mining & U & 96 & 1,086,404 \\
\cite{konect:slashdot-threads} & \textsf{SD} & Slashdot threads & U M & 51,083 & 140,778 \\
\cite{konect:Rocha2010} & \textsf{SX} & Sexual escorts & B & 16,730 & 50,632 \\
\cite{b554} & \textsf{TO} & Internet topology & U & 34,761 & 171,403 \\
\cite{konect:opsahl09} & \textsf{UC} & UC Irvine messages & U M & 1,899 & 59,835 \\
\cite{konect:opsahl10} & \textsf{UF} & UC Irvine forum & B & 1,421 & 33,720 \\

      \bottomrule
      \multicolumn{6}{l}{
        U Unipartite network
      } \\
      \multicolumn{6}{l}{
        B Bipartite network
      } \\
      \multicolumn{6}{l}{
        M Network with multiple edges
      }
    \end{tabular}
  }
\end{table}

There are two ways in which the evolution of a network can be measured,
which we both perform in our experiments:
\begin{description}
\item \textsc{Full}:
  The first type of measurement
  looks at the evolution of the complete network for times ranging from
  the network's inception to the last added edge. 
  In this type of experiment, the number of nodes $V$ in
  the network varies, as new nodes appear in the network.
  The drawback of this method is that a network is in general not
  connected, and some measures, such as the diameter, can only be
  computed for a connected network.  Thus, a method is needed to study
  only the connected network. 
\item \textsc{Connected}:
  In the second type of measurement, the set of vertices $V_1$ is fixed
  at some time $t_1$, the largest connected component of nodes $\bar
  V_1$ is found, and then only the subnetworks consisting of the nodes in
  $\bar V_1$ are considered at later times $t > t_1$.  
  This ensures that the network is connected, but restricts the
  experiments to times after the time $t_1$. 
  This method is also used for evaluating the problem of link
  prediction, under the assumption that links connecting two
  disconnected components cannot be predicted sensibly. This method is
  then suitable because it ensures that each new edge connects two nodes
  already connected. 
\end{description}

Let $\{G_i\}_{i=1}^N$ be a set of $N$ network datasets.
Each network is split into one hundred timepoints $t=1, \dotsc, 100$,
each containing $\lfloor |E| t/100 \rfloor$ of the oldest of all edges
$E$ in the network. Let $G_i^t$ be the network $G_i$ containing all
edges up to time $t$.  Then each measure $f$ is computed for all networks
at all timepoints.  Thus,
$f(G_i^1), f(G_i^2), \dotsc, f(G_i^{100})$
is the time series representing the evolution of the network measure $f$
for the \textsc{Full} variant.  

For the \textsc{Connected} variant, we chose the starting time to be $t_1=75$,
i.e., three quarters of the available time range. This specific choice
is arbitrary, and constitutes a trade-off between the requirement that the chosen time must not be too early,
as otherwise the connected component is too small, and the requirement that the chosen time
must not be too late, as otherwise the considered time range is too short.  

To test whether a single network dataset has a shrinking diversity for
a given diversity measure, 
we apply the Mann--Kendall test~\cite{b744}. 
Given a series $(x_i)$, the Mann--Kendall test consists of applying a
$t$-test to all pairwise differences $x_i - x_j$.  
We accept the hypothesis of a decreasing diversity for one
network/measure combination when the $p$-value is below the threshold of
$\alpha=0.05$. 
The result of the individual Mann--Kendall tests for all network/measure
combinations are shown in
Appendix~I. 
%% the supplementary material to this article. 
For simplicity, we will describe the test procedure for measures of
diversity such as the entropy.  For measures that measure the opposite
of diversity, the test is analogous. 
For each measure $f$, we aggregate the results for all networks and
test the hypothesis that the diversity measure is decreasing.  The null hypothesis
is thus that the measure $f$ is not decreasing. Since the Mann--Kendall
test is performed to a value of $\alpha=0.05$, the probability of
having $k$ out of $n$ successes for the measure $f$ equals a binomial
distribution with probability parameter $\alpha$. Thus, the $p$-value
for the null hypothesis that the measure is not decreasing equals 
$p = \sum_{x=k}^n {k \choose n} \alpha^k (1 - \alpha)^{n-k}$. 
We compute this $p$-value for each measure $f$ and accept the hypothesis
that the diversity measure $f$ in decreasing when $p < \alpha = 0.05$. 

The summarized results of the statistical tests are shown in
Table~\ref{tab:test}. 

\begin{table}[t]
  \caption{
    \label{tab:test}
    Statistical significance tests for shrinking of diversity according
    to the eleven different measures. 
    Statistically significant trends are shown as \emph{Up} and
    \emph{Down}. No statistically significant trend is denoted by a dash
    (---).  
    Numbers in parentheses give the number of networks following the
    given trend according to the Mann--Kendall test, out of all 27
    networks.  (In cases without a trend, the number counts the networks
    following the predicted trend.)
  }
  \makebox[\textwidth]{
    \centering
    \scalebox{0.9}{
  \begin{tabular}{c|c|ll|l|l}
      \toprule
      & \textbf{Measure} & \multicolumn{2}{c|}{\textbf{Observed trends}} &
      \textbf{Predicted trends} & \multicolumn{1}{|c}{\textbf{Monotonicity}} \\
      & & \textsc{Full} & \textsc{Connected} & & \textsc{Connected} \\ 
      \midrule
      & $d$			& (24) Up     & (27) Up      &  & Up \\
      \midrule
      \multirow{4}{*}{\begin{sideways}\textbf{Pref. att.}\end{sideways}} 
      & $G$                       & (24) Up  & (17) ---  & Up   &      \\
      & $J$                       & (23) Up  & (20) Up   & Down &      \\
      & $\gamma$	              & (21) Down & (25) Down & Down &      \\
      & $H_{\mathrm{er}}$            & (19) Down & (12) ---  & Down &     \\	
      \midrule
      \multirow{4}{*}{\begin{sideways}\scalebox{0.8}[1]{\textbf{Connect.}}\end{sideways}} 
      & $\delta_{0.9}$             & (18) Down & (26) Down & Down & Down \\
      & $\vartheta_r(n)$          & (10) ---  & (22) Down & Down &      \\
      & $C_{\mathrm r}$             & (12) --- & (22) Down & Down & Down \\
      & $a$                       & (15) --- & (27) Up  & Up & Up   \\
      \midrule
      \multirow{3}{*}{\begin{sideways}\scalebox{0.8}[1]{\textbf{L. pred.}}\end{sideways}} 
      & $c$			      & (\phantom{0}7) --- $^a$  & (10) Up $^a$   & Up & \\
      & $\mathrm{rank}_{\mathrm F}$ & (13) --- & (19) Down & Down &      \\
      & $\alpha$                  & (19) Up  & (23) Up   & Up   &      \\
      \bottomrule
  \end{tabular}
  }
  }
      $^a$
      For the clustering coefficient, the total number of networks
      is~13, since bipartite networks are excluded.  
\end{table}

\subsection{Discussion}
Out of the eleven measures of structural diversity, all but one show a
temporal trend consistent with shrinking in either the \textsc{Full} or
\textsc{Connected} case.  
Two measures show shrinking diversity in both
cases:  the power-law exponent and the diameter. Three
measures are predicted to show shrinking diversity in the 
\textsc{Connected} case mathematically, and also do so in the
experiments. 

Our experimental results thus show that for a large majority of
structural network measures, a trend exists that can be interpreted as
shrinking diversity.  This leads us to conclude that the notion of
structural diversity is a legitimate one, which explains in a more
intuitive way the temporal evolution of different network measures.
Thus, the shrinking diversity hypothesis gives an additional
justification for models of preferential attachment, connectivity and
link prediction.  Detailed discussions follow. 

\paragraph{Preferential Attachment}
Out of the four measures of degree equality or inequality, three show
statistically significant trends consistent with shrinking diversity.
The single exception is Jain's index $J$, which shows a trend consistent
with increasing diversity. 
The preferential attachment model is thus validated by our
experiments, up to the differing behavior of Jain's index. 
For the fractional rank $\mathrm{rank}_{\mathrm F}$, which can be
interpreted as following from a process of preferential attachment on
eigenvectors, we observe a shrinking trend in the connected case. 
The different behavior of Jain's index is intriguing.  On the face of
it, we would be inclined to conclude that the preferential attachment
hypothesis is not correct, according to our experiments with Jain's
index. However, the clear and consistent results for the Gini
coefficient, power-law 
exponent and entropy lead us rather to conclude that Jain's index is not
a typical measure of diversity, and correlates negatively with other
such measures.  

\paragraph{Connectivity}
Of the three types of network measures studied in this article, the
measures based on connectivity are the weakest in matters of shrinking
diversity.  For three of them, a 
proof exists that they must evolve according to a shrinking diversity in
the connected case.  Thus, the only nontrivial result is the shrinking
diameter, which is observed even when taking new nodes into account,
and the shrinking random walk return probability in the connected case. 

The diameter is decreasing in general.  We observe however 
that the diameter varies from one timepoint to the other sometimes as
much as its overall trend. 
This is an indication that the diameter is not a robust measure.  This
is in opposition to reference~\cite{b242}, where a consistently
shrinking effective diameter is reported for multiple networks.

The random walk return probability $\vartheta_r(n)$ is shrinking for
most networks, in accordance with a shrinking diversity. This result
also implies that the eigenvalues of the normalized Laplacian matrix
$\mathbf Z$ move towards the value one, or equivalently, that the
eigenvalues of the normalized adjacency matrix $\mathbf D^{-1/2} \mathbf
A \mathbf D^{-1/2} = \mathbf I - \mathbf Z$ shrink. 

The relative controllability $C_{\mathrm r}$ is decreasing for almost all
networks.  This pattern is more consistent than for the three other
connectivity measures.  A decreasing relative number of driver nodes means that less and less
vertices are necessary to be controlled in order to control the whole
network.  Thus, the diversity of the network is going down.  

\paragraph{Link Prediction}
All three network measures based on link prediction evolve in a way
consistent with shrinking diversity in the connected case, but only the
power-law exponent does so in the unconnected case. This result confirms
that link prediction methods can normally only be applied to connected
networks, and that they do not give sensible results for unconnected
nodes. For instance, a neighborhood-based link prediction method 
cannot predict a new edge connecting two disconnected components, since
they always have zero neighbors in common. 

The fractional rank is decreasing for the majority of networks, but by
far not for all networks. This is an indication that common graph
kernels such as the matrix exponential and the Neumann kernel are
accurate link prediction functions. 
By interpreting each latent dimension of the eigenvalue decomposition of
$\mathbf A$ as a community or topic (depending on the network) and the
corresponding eigenvalue as the weight of the community or topic, 
implies that large communities or topics get larger over time, and go on
to dominate smaller topics. 

The clustering coefficient is the only measure considered that is only
meaningful for unipartite networks, of which there are thirteen in our
tests. Despite this reduced number, the clustering coefficient is
increasing in ten of these networks, showing that the \emph{triangle
  closing} model is correct in a majority of networks. 

\section{Conclusion}
\label{sec:conclusion}
The evolution of networks can indeed be understood in terms of a
shrinking diversity, and common models of network evolution thus admit
an interpretation in terms of diversity.  The preferential attachment
model is thus true because it predicts that new edges attach to popular
nodes, decreasing the diversity of connections.  The connectivity
interpretation implies that connectivity is generally increasing over
time in real-world networks, an observation in line with previous
results.  Finally, link prediction algorithms that are found to perform
well in practice are justified as they presuppose shrinking diversity.  
These results justify the notion of structural diversity, and show that
it is a primary driver in network evolution, and thus represents a basis
for many temporal network analysis methods.  

In terms of overall diversity, which may also include non-structural
measures, our methods can however only give answer to the point that
diversity can be represented equivalently as a network.  While for
instance the diversity of movies watched by the public is well
represented by the structural diversity of the bipartite person--movie
network, other types of diversity may not.  The ubiquity of networks as
a model however suggests that this is only rarely the case:  many things
whose diversity we are interested in such as opinions, languages,
friendships, words, etc., can be represented as nodes in a network, and
are thus amenable to our methodology. 
The positive results in our study should of course not be taken for
natural; network evolution rules that defy the shrinking diversity
hypothesis can of course not be ruled out by it, and may very well give
particularly salient insight into processes at work in networks. 

\paragraph{Acknowledgments}
We thank Sergej Sizov, Julia Perl, Damien Fay and Felix Schwagereit.  
The research leading to these results has received funding from the
European Community's Seventh Framework Programme under grant agreement
n\textsuperscript{o}~257859, ROBUST. 

\let\oldbibliography\thebibliography
\renewcommand{\thebibliography}[1]{%
  \oldbibliography{#1}%
  \setlength{\itemsep}{-6.00pt}%
}
\bibliographystyle{abbrv}
\bibliography{ref,kunegis,konect,diversity} 

\begin{thebibliography}{10}

\bibitem{aldrich1976}
H.~Aldrich.
\newblock Resource dependence and interorganizational relations.
\newblock {\em Administration and Society}, 7(4):419--454, 1976.

\bibitem{b742}
A.~B. Atkinson.
\newblock On the measurement of inequality.
\newblock {\em J. of Economic Theory}, 2(3):244--263, 1970.

\bibitem{b439}
A.-L. Barabási and R.~Albert.
\newblock Emergence of scaling in random networks.
\newblock {\em Science}, 286(5439):509--512, 1999.

\bibitem{b771}
A.-L. Barabási, H.~Jeong, Z.~Neda, E.~Ravasz, and A.~Schubert.
\newblock Evolution of the social network of scientific collaborations.
\newblock {\em Physica A}, 311(3--4):590--614, 2002.

\bibitem{b116}
B.~Bollobás.
\newblock {\em Modern Graph Theory}.
\newblock Springer, 1998.

\bibitem{b748}
J.~A. Bondy and U.~S.~R. Murty.
\newblock {\em Graph Theory with Applications}.
\newblock American Elsevier Pub. Co, 1976.

\bibitem{intersectoral}
L.~D. Brown and A.~D.
\newblock Participation, social capital, and intersectoral problem solving:
  {African} and {Asian} cases.
\newblock {\em World Development}, 24(9):1467--1479, 1996.

\bibitem{children-television}
J.~A. Bryant and P.~R. Monge.
\newblock The evolution of the children's television community.
\newblock {\em Int. J. of Communication}, 2:160--192, 2008.

\bibitem{b508}
J.~Bunch, C.~Nielsen, and D.~Sorensen.
\newblock Rank-one modification of the symmetric eigenproblem.
\newblock {\em Numerische Math.}, 31(1):31--48, 1978.

\bibitem{b532}
A.~Chaintreau, P.~Hui, J.~Crowcroft, C.~Diot, R.~Gass, and J.~Scott.
\newblock Impact of human mobility on opportunistic forwarding algorithms.
\newblock {\em IEEE Trans. on Mobile Computing}, 6(6):606--620, 2007.

\bibitem{b565}
M.~D. Choudhury, H.~Sundaram, A.~John, and D.~D. Seligmann.
\newblock Social synchrony: Predicting mimicry of user actions in online social
  media.
\newblock In {\em Proc. Int. Conf. on Computational Science and Engineering},
  pages 151--158, 2009.

\bibitem{b702}
P.~B. Coulter.
\newblock {\em Measuring Inequality: A Methodological Handbook}.
\newblock Westview Press, 1989.

\bibitem{inter3}
M.~L. Doerfel and M.~Haseki.
\newblock Networks, disrupted: Media use as an organizing mechanism for
  rebuilding.
\newblock {\em New Media \& Society}, page 1461444813505362, 2013.

\bibitem{b665}
N.~Eagle and A.~S. Pentland.
\newblock {Reality} {Mining}: Sensing complex social systems.
\newblock {\em Personal Ubiquitous Computing}, 10(4):255--268, 2006.

\bibitem{b535}
M.~Faloutsos, P.~Faloutsos, and C.~Faloutsos.
\newblock On power-law relationships of the {Internet} topology.
\newblock {\em SIGCOMM Computer Commun. Rev.}, 29:251--262, 1999.

\bibitem{b425}
D.~Fay, H.~Haddadi, A.~Thomason, A.~W. Moore, R.~Mortier, A.~Jamakovic,
  S.~Uhlig, and M.~Rio.
\newblock Weighted spectral distribution for {Internet} topology analysis:
  Theory and applications.
\newblock {\em IEEE Trans. on Networking}, 18(1):164--176, 2010.

\bibitem{b652}
M.~Fiedler.
\newblock Algebraic connectivity of graphs.
\newblock {\em Czechoslovak Math. J.}, 23(98):298--305, 1973.

\bibitem{b320}
C.~Gkantsidis, M.~Mihail, and E.~Zegura.
\newblock Spectral analysis of {Internet} topologies.
\newblock In {\em Proc. Joint Conf. IEEE Computer and Communications
  Societies}, pages 364--374, 2003.

\bibitem{b537}
K.-I. Goh, B.~Kahng, and D.~Kim.
\newblock Spectra and eigenvectors of scale-free networks.
\newblock {\em Phys. Rev. E}, 64(5):051903, 2001.

\bibitem{www.grouplens.org/node/73}
{GroupLens Research}.
\newblock {MovieLens} data sets.
\newblock \url{http://www.grouplens.org/node/73}, October 2006.

\bibitem{konect:slashdot-threads}
V.~Gómez, A.~Kaltenbrunner, and V.~López.
\newblock Statistical analysis of the social network and discussion threads in
  {Slashdot}.
\newblock In {\em Proc. Int. World Wide Web Conf.}, pages 645--654, 2008.

\bibitem{hannan1977}
M.~T. Hannan and J.~Freeman.
\newblock The population ecology of organizations.
\newblock {\em The American J. of Sociology}, 82:929--964, 1977.

\bibitem{hannan1984}
M.~T. Hannan and J.~Freeman.
\newblock Structural inertia and organizational change.
\newblock {\em American Sociological Review}, 49(2):149--164, 1984.

\bibitem{hannan1988}
M.~T. Hannan and J.~Freeman.
\newblock The ecology of organizational mortality: {American} labor unions.
\newblock {\em The American J. of Sociology}, 94(1):25--52, 1988.

\bibitem{hardin1960}
G.~Hardin.
\newblock The competitive exclusion principle.
\newblock {\em Science}, 131(3409):1292--1297, April 1960.

\bibitem{hawley1981}
A.~H. Hawley.
\newblock Human ecology: Persistence and change.
\newblock {\em American Behavioral Scientist}, 24(3):423--444, 1981.

\bibitem{hawley1986}
A.~H. Hawley.
\newblock {\em Human Ecology: A Theoretical Essay}.
\newblock University of Chicago Press, 1986.

\bibitem{group-bias}
M.~Hewstone, M.~Rubin, and H.~Willis.
\newblock Intergroup bias.
\newblock {\em Annu. Rev. Psychol.}, 53:575--604, 2002.

\bibitem{hinds2000}
P.~J. Hinds, K.~M. Carley, and D.~Krackhardt.
\newblock Choosing work group members: Balancing similarity, competence, and
  familiarity.
\newblock {\em Organizational Behavior and Human Decision Processes},
  81(2):226--251, 2000.

\bibitem{konect:sociopatterns}
L.~Isella, J.~Stehlé, A.~Barrat, C.~Cattuto, J.-F. Pinton, and W.~V. den
  Broeck.
\newblock What's in a crowd? analysis of face-to-face behavioral networks.
\newblock {\em J. of Theoretical Biology}, 271(1):166--180, 2011.

\bibitem{b137}
T.~Ito, M.~Shimbo, T.~Kudo, and Y.~Matsumoto.
\newblock Application of kernels to link analysis.
\newblock In {\em Proc. Int. Conf. on Knowledge Discovery in Data Mining},
  pages 586--592, 2005.

\bibitem{b740}
R.~Jain, D.~Chiu, and W.~Hawe.
\newblock A quantitative measure of fairness and discrimination for resource
  allocation in shared computer system.
\newblock Technical Report 301, DEC Research, 1984.

\bibitem{b263}
J.~Kandola, J.~Shawe-Taylor, and N.~Cristianini.
\newblock Learning semantic similarity.
\newblock In {\em Advances in Neural Information Processing Systems}, pages
  657--664, 2002.

\bibitem{b345}
B.~Klimt and Y.~Yang.
\newblock The {Enron} corpus: A new dataset for email classification research.
\newblock In {\em Proc. Eur. Conf. on Machine Learning}, pages 217--226, 2004.

\bibitem{b156}
R.~Kondor and J.~Lafferty.
\newblock Diffusion kernels on graphs and other discrete structures.
\newblock In {\em Proc. Int. Conf. on Machine Learning}, pages 315--322, 2002.

\bibitem{b700}
R.~Kumar, J.~Novak, and A.~Tomkins.
\newblock Structure and evolution of online social networks.
\newblock In {\em Proc. Int. Conf. on Knowledge Discovery and Data Mining},
  pages 611--617, 2006.

\bibitem{konect}
J.~Kunegis.
\newblock {KONECT} -- {The} {Koblenz} {Network} {Collection}.
\newblock In {\em Proc. Int. Conf. on World Wide Web Companion}, pages
  1343--1350, 2013.

\bibitem{kunegis:spectral-network-evolution}
J.~Kunegis, D.~Fay, and C.~Bauckhage.
\newblock Network growth and the spectral evolution model.
\newblock In {\em Proc. Int. Conf. on Information and Knowledge Management},
  pages 739--748, 2010.

\bibitem{kunegis:spectral-evolution}
J.~Kunegis, D.~Fay, and C.~Bauckhage.
\newblock Spectral evolution in dynamic networks.
\newblock {\em Knowledge and Information Systems}, 37(1):1--36, 2013.

\bibitem{kunegis:spectral-transformation}
J.~Kunegis and A.~Lommatzsch.
\newblock Learning spectral graph transformations for link prediction.
\newblock In {\em Proc. Int. Conf. on Machine Learning}, pages 561--568, 2009.

\bibitem{kunegis:power-law}
J.~Kunegis and J.~Preusse.
\newblock Fairness on the web: Alternatives to the power law.
\newblock In {\em Proc. Web Science Conf.}, pages 175--184, 2012.

\bibitem{kunegis:network-rank}
J.~Kunegis, S.~Sizov, F.~Schwagereit, and D.~Fay.
\newblock Diversity dynamics in online networks.
\newblock In {\em Proc. Conf. on Hypertext and Social Media}, pages 255--264,
  2012.

\bibitem{b461}
J.~Leskovec, L.~Backstrom, R.~Kumar, and A.~Tomkins.
\newblock Microscopic evolution of social networks.
\newblock In {\em Proc. Int. Conf. on Knowledge Discovery and Data Mining},
  pages 462--470, 2008.

\bibitem{b551}
J.~Leskovec, D.~Huttenlocher, and J.~Kleinberg.
\newblock Governance in social media: A case study of the {Wikipedia} promotion
  process.
\newblock In {\em Proc. Int. Conf. on Weblogs and Social Media}, pages 98--105,
  2010.

\bibitem{b242}
J.~Leskovec, J.~Kleinberg, and C.~Faloutsos.
\newblock Graph evolution: Densification and shrinking diameters.
\newblock {\em ACM Trans. Knowledge Discovery from Data}, 1(1):1--40, 2007.

\bibitem{b673}
Y.-Y. Liu, J.-J. Slotine, and A.-L. Barabási.
\newblock Controllability of complex networks.
\newblock {\em Nature}, 473:167--173, May 2011.

\bibitem{b744}
H.~B. Mann.
\newblock Nonparametric tests against trend.
\newblock {\em Econometrica}, 13(3):245--259, 1945.

\bibitem{b367}
P.~Massa and P.~Avesani.
\newblock Controversial users demand local trust metrics: an experimental study
  on {epinions.com} community.
\newblock In {\em Proc. American Association for Artificial Intelligence
  Conf.}, pages 121--126, 2005.

\bibitem{b672}
M.~Mihail and C.~H. Papadimitriou.
\newblock On the eigenvalue power law.
\newblock In {\em Proc. Int. Workshop on Randomization and Approximation
  Techniques}, pages 254--262, 2002.

\bibitem{monge2003}
P.~R. Monge and N.~S. Contractor.
\newblock {\em Theories of Communication Networks}.
\newblock Oxford University Press, 2003.

\bibitem{monge2008}
P.~R. Monge, B.~M. Heiss, and D.~B. Margolin.
\newblock Communication network evolution in organizational communities.
\newblock {\em Communication Theory}, 18(4):449--477, 2008.

\bibitem{b408}
M.~E.~J. Newman.
\newblock Power laws, {Pareto} distributions and {Zipf}'s law.
\newblock {\em Contemporary Phys.}, 46(5):323--351, 2006.

\bibitem{konect:opsahl09}
T.~Opsahl and P.~Panzarasa.
\newblock Clustering in weighted networks.
\newblock {\em Social Networks}, 31(2):155--163, 2009.

\bibitem{konect:opsahl10}
T.~Opsahl and P.~Panzarasa.
\newblock Triadic closure in two-mode networks: Redefining the global and local
  clustering coefficients.
\newblock {\em Social Networks}, 34, 2011.

\bibitem{konect:Rocha2010}
L.~E.~C. Rocha, F.~Liljeros, and P.~Holme.
\newblock Information dynamics shape the sexual networks of {Internet}-mediated
  prostitution.
\newblock {\em Proc. of the National Academy of Sciences}, 107(13):5706--5711,
  2010.

\bibitem{diffusion-of-innovations}
E.~Rogers.
\newblock {\em Diffusion of Innovations}.
\newblock Simon and Schuster, 1962.

\bibitem{said:social-similarity}
A.~Said, E.~W. De~Luca, and {\,{S}}.~Albayrak.
\newblock How social relationships affect user similarities.
\newblock In {\em Proc. IUI Workshop on Social Recommender Systems}, 2010.

\bibitem{inter1}
G.~D. Saxton and M.~A. Benson.
\newblock Social capital and the growth of the nonprofit sector.
\newblock {\em Social Science Quarterly}, 86(1):16--35, 2005.

\bibitem{inter2}
M.~Shumate and A.~O'Connor.
\newblock The symbiotic sustainability model: Conceptualizing ngo--corporate
  alliance communication.
\newblock {\em J. of Communication}, 60:577--609, 2010.

\bibitem{ingroups-outgroups}
H.~Tajfel, M.~G. Billig, R.~P. Bundy, and C.~Flament.
\newblock Social categorization and intergroup behaviour.
\newblock {\em Eur. J. of Social Psychology}, 1(2):149--178, 1971.

\bibitem{b741}
H.~Theil.
\newblock The information approach to demand analysis.
\newblock {\em Econometrica}, 33(1):67--87, 1965.

\bibitem{b480}
B.~Viswanath, A.~Mislove, M.~Cha, and K.~P. Gummadi.
\newblock On the evolution of user interaction in {Facebook}.
\newblock In {\em Proc. Workshop on Online Social Networks}, pages 37--42,
  2009.

\bibitem{b693}
B.~Wang, H.~Tang, C.~Guo, and Z.~Xiu.
\newblock Entropy optimization of scale-free networks robustness to random
  failures.
\newblock {\em Physica A}, 363(2):591--596, 2006.

\bibitem{b228}
D.~J. Watts and S.~H. Strogatz.
\newblock Collective dynamics of `small-world' networks.
\newblock {\em Nature}, 393(1):440--442, 1998.

\bibitem{download.wikimedia.org}
{Wikimedia Foundation}.
\newblock Wikimedia downloads.
\newblock \url{http://dumps.wikimedia.org/}, January 2010.

\bibitem{b663}
J.~H. Wilkinson.
\newblock {\em The Algebraic Eigenvalue Problem}.
\newblock Oxford University Press, 1965.

\bibitem{b681}
J.~Wu, Y.-J. Tan, H.-Z. Deng, and D.-Z. Zhu.
\newblock A new measure of heterogeneity of complex networks based on degree
  sequence.
\newblock In {\em Unifying Themes in Complex Systems}, pages 66--73. 2010.

\bibitem{b554}
B.~Zhang, R.~Liu, D.~Massey, and L.~Zhang.
\newblock Collecting the {Internet} {AS}-level topology.
\newblock {\em SIGCOMM Computer Commun. Rev.}, 35(1):53--61, 2005.

\end{thebibliography}

\section{Appendix I:  Network Diversity Results}
Figure~\ref{fig:evol} shows the evolution of all eleven network diversity
measures and the average degree $d$ applied to our collection of network
datasets. The plots show 
the \textsc{Full} scenario, i.e., the evolution from $t=1$ to $t=100$
including all vertices and edges. 

\newcommand{\wEvol}{0.061\textwidth}
\newcommand{\tSpace}{\,}
\begin{figure}
  \renewcommand{\arraystretch}{0.18}
  \centering
  \scalebox{0.8}{
  \begin{tabular}{c@{\;}c@{\tSpace}c@{\;}c@{\tSpace}c@{\tSpace}c@{\tSpace}c@{\tSpace}c@{\tSpace}c@{\tSpace}c@{\tSpace}c@{\tSpace}c@{\tSpace}c@{\tSpace}c}
    & $d$ % density
    & $G$ % gini 
    & $J$ % jain
    & $\gamma$ % power
    & $H_{\mathrm{er}}$ % dentropyn
    & $\delta_{0.9}$ % diameter 
    & $\vartheta_r(n)$ % network rank 4 
    & $C_{\mathrm r}$ % (relative) controllability 
    & $a$ % alcon 
    & $c$ % clusco
    & $\mathrm{rank}_{\mathrm F}$ % network_rank_sq 
    & $\alpha$ % epower
    \\
    \input{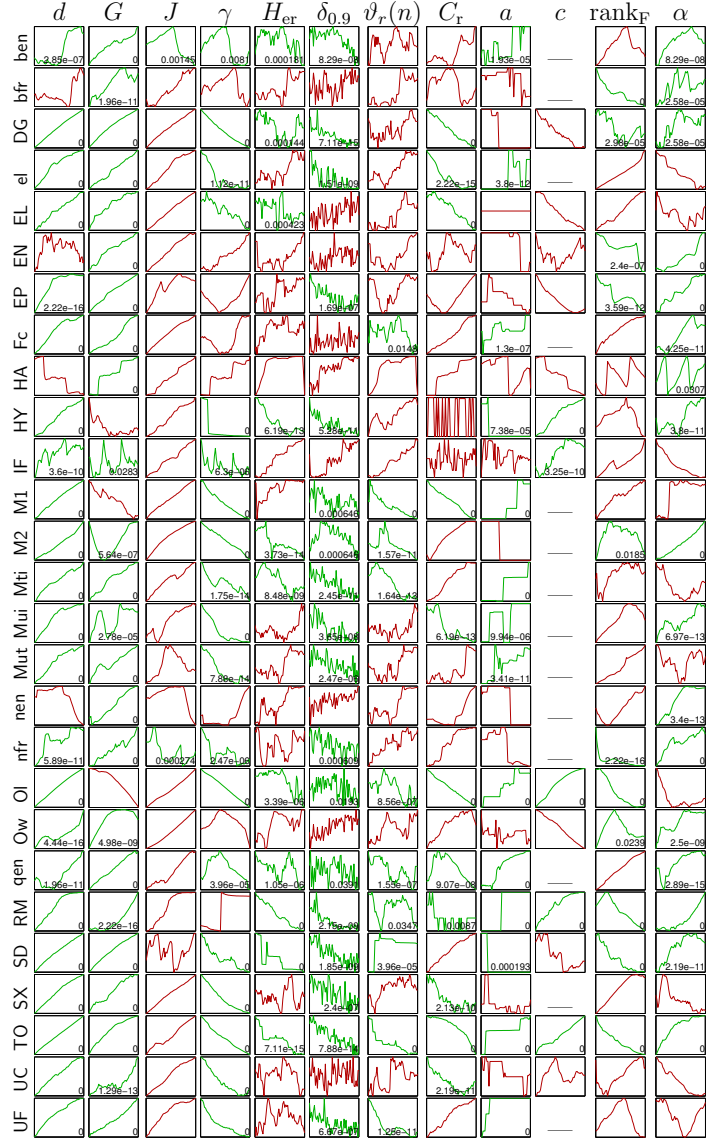}
  \end{tabular}
  }
  \caption{
    \label{fig:evol}
    The evolution of all eleven diversity measures for all evaluated network
    datasets. 
    The plots correspond to the \textsc{Full} scenario, i.e., time on
    the X axis goes from $t=1$ to $t=100$ and all vertices and edges are
    included. 
    The color of the plot indicates whether the measure is evolving
    according to the prediction (green, $p$-value shown) or not (red, no
    $p$-value shown).  
  }
\end{figure}

\end{document}